\documentclass[twocolumn,prl,aps,superscriptaddress,english,floatfix]{revtex4-1}
\usepackage{amsmath}
\usepackage{amssymb}
\usepackage{amsfonts}
\usepackage[pdftex]{graphicx}
\usepackage{subfigure} 
\usepackage[english]{babel}
\usepackage{braket}
\usepackage{color}
\usepackage{mathtools}
\usepackage{threeparttable}
\usepackage[colorlinks,linkcolor=blue,anchorcolor=blue,citecolor=blue,urlcolor=blue]{hyperref}
\usepackage{babel}

\makeatother

\begin{document}

\preprint{APS/123-QED}

\title{Quantum Critical Detector: Amplifying Weak Signals Using First-Order Dynamical Quantum Phase Transitions}

\author{Li-Ping Yang}

\affiliation{Birck Nanotechnology Center and Purdue Quantum Center,School of Electrical
and Computer Engineering, Purdue University, West Lafayette, IN 47906,
U.S.A.}

\author{Zubin Jacob}
\email{zjacob@purdue.edu}

\homepage{http://www.zjresearchgroup.org/}

\affiliation{Birck Nanotechnology Center and Purdue Quantum Center,School of Electrical
and Computer Engineering, Purdue University, West Lafayette, IN 47906,
U.S.A.}
\begin{abstract}
 We introduce a first-order quantum-phase-transition model, which exhibits giant sensitivity $\chi \propto N^2$ at the critical point. Exploiting this effect, we propose a quantum critical detector (QCD) to amplify weak input signals. The time-dynamic QCD functions by triggering a first-order dynamical quantum phase transition in a system of spins with long-range interactions coupled to a bosonic mode. We numerically demonstrate features of the dynamical quantum phase transition, which leads to a time-dependent quantum gain. We also show the linear scaling  with the spin number $N$ in both the quantum gain and the corresponding signal-to-quantum noise ratio during the time evolution of the device. Our QCD can be a resource for metrology,  weak signal amplification, and single photon detection.

\end{abstract}
\maketitle
Amplification is the key process for weak signal detection. There exist two main quantum amplification schemes. In the first kind, the weak input  signal is directly amplified to generate a large output signal, as in quantum linear amplifiers~\cite{caves1982quantum,louisell1961quantum,Mollow1967parametric1,Gavish2004generalized}. It is widely utilized in circuit QED for parametric amplification in Josephson junction amplifiers~\cite{roy2016introduction}. Another scheme of amplification exists where  the weak input perturbation functions as a control signal of an optimally biased device near the critical point. The practical realizations include single-electron transistors~\cite{devoret2000amplifying} and single-photon detectors~\cite{eisaman2011invited}. An important distinction in the critically biased amplifiers from quantum linear amplifiers is that the input and output information carriers can be fundamentally different (eg: input photons and output electrons). The goal of this paper is to propose a class of biased detectors  with an amplification scheme that exploits quantum criticality in first-order dynamical quantum phase transitions (DQPT). Our proposed device can be considered as an engineered quantum analog of widely utilized detectors which exploit naturally occurring thermodynamic phase transitions (eg: superconducting to normal metal~\cite{gol2001picosecond}).

Quantum phase transition (QPT) at zero temperature describes an abrupt change in the ground state of a many-body system~\cite{sachdev2007quantum,quan2006decay}. Most of the QPTs discovered in physical systems are of second-order~\cite{lieb1961two,lipkin1965validity,meshkov1965validity,glick1965validity,hepp1973superradiant,wang1973phase} and they have also been proposed as a resource for metrology. However, no significant change in the number of excitations or energy transfer between subsystems occurs during a second-order QPT. To obtain an observable change in the output, a large parameter  variation is required which can not be induced by a weak input signal. Thus, high quantum gain can not be obtained using traditional second-order QPTs limiting their applicability for weak signal amplification. On the contrary, in a first-order QPT, even a very small parameter variation at the critical point can lead to a significantly large change in the values of  physical observables. Therefore, a universal model exhibiting first-order QPT with a well-understood microscopic mechanism can provide a natural platform for quantum amplification~\cite{caves1982quantum}, quantum metrology~\cite{giovannetti2004quantum,giovannetti2006quantum}, and lead to new types of single-photon detectors.

We emphasize that time dynamics near the critical point fundamentally determines whether QPTs can be a practical resource for amplification and detection. Specifically, critical scaling in QPT results only from the transition between the ground states of two different phases but can not be realized by a unitary adiabatic evolution. Thus, a dynamical detection event, which does not connect these two ground states, may have totally different critical scaling. Few first-order QPT models have already been found~\cite{lee2004first,Ovchinnikov2003anti-ferro,vidal2004entanglement,del2016nonequilium}, but the dynamics of these QPTs around the critical point have not been revealed. The application of these first-order QPTs is therefore an open problem, as practical detection events and amplifications are fundamentally dynamical processes.

\begin{figure*}
\includegraphics[width=12cm]{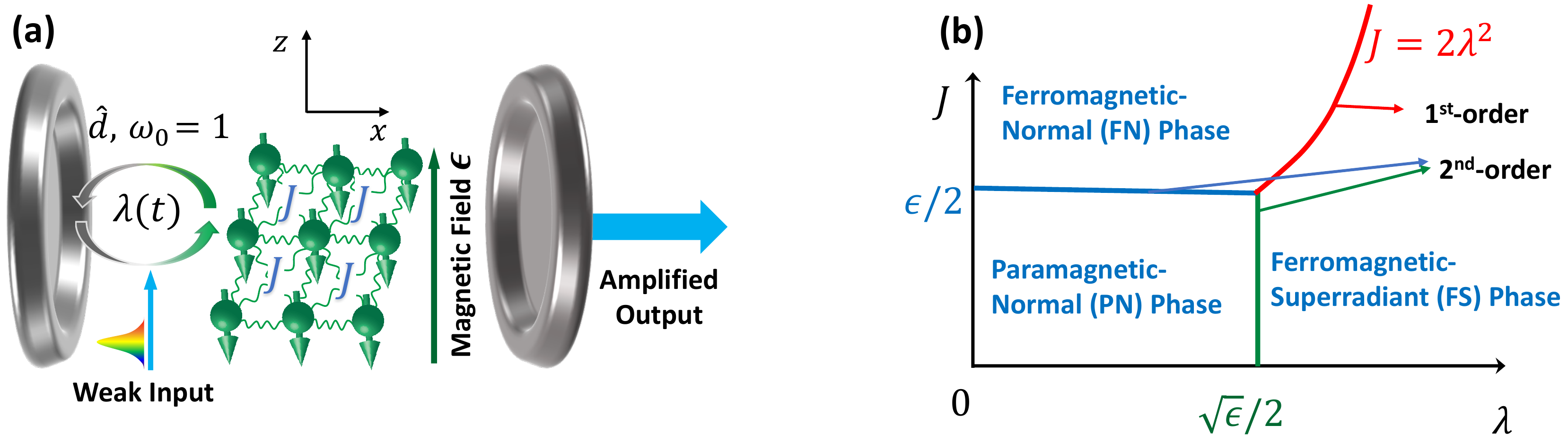}
\caption{\label{fig:schematic} (a) Schematic of the quantum critical detector (QCD). The bosonic mode (resonant cavity $\hat{d}$-mode) with frequency $\omega_{0}=1$ is the output mode. The spins are immersed in a homogeneous magnetic field along $z$-axis inducing an energy splitting $\epsilon$. The spin-boson coupling $\lambda$ is in $x$-direction and the all-to-all spin-spin coupling $J$ is along $y$-axis. The input weak signal leads to a small time-dependent variation in spin-boson coupling $\lambda(t)$ and triggers a first-order dynamical quantum phase transition if the system is optimally biased around the critical point. The energy pre-stored in the spins  transfers to the bosonic mode and realize the amplification in our QCD. (b) Phase diagram of our model. The green, blue, and red lines give the boundaries
of the three quantum phases. In the strong-coupling regime $J>J_{c,{\rm II}}\equiv\epsilon/2$
and $\lambda>\lambda_{c,{\rm II}}\equiv\sqrt{\epsilon}/2$, the QPT between the FN phase and the FS phase (crossing the red line) is of first order. The other two QPTs are of second-order.}
\end{figure*}

In this letter, we introduce a first-order QPT model, composed of a bosonic mode and a spin ensemble with long-range interaction. We numerically show that there exists a critical point in this system where  the sensitivity $\chi$ diverges with $N^2$-scaling (N the spin number). This scaling is much faster than previous first-order phase transitions~\cite{gammelmark2011phase,raghunandan2018high} and provides extraordinary high sensitivity for quantum metrology. This first-order QPT and the giant sensitivity in our model fundamentally originate from the competition between two phases with long-range spin order. Exploiting this unique criticality, we propose a quantum critical detector (QCD) utilizing a weak input signal triggered first-order DQPT [see the schematic in Fig.~\ref{fig:schematic} (a)]. Via direct numerical evaluations, we demonstrate the time-dynamical features of a QCD consisting of $80$ spins interacting with a bosonic mode. We show the existence of a first-order DQPT in our QCD which sheds light on the microscopic origin of time-dependent quantum gain $g(t)$. Finally, we show the linear scaling in both the maximum quantum gain and the corresponding signal-to-quantum noise ratio (SQNR) during the time-evolution of our detector revealing high figures of merit. 

\textit{First-Order Quantum Phase Transition}\textemdash The key element of a QCD is the first-order QPT based quantum amplification. Here, we introduce a specific first-order QPT model composed of a bosonic mode and an interacting spin ensemble,
\begin{equation}
H=\hat{d}^{\dagger}\hat{d}\!+\!\frac{\lambda}{\sqrt{N}}\!\sum_{j=1}^{N}\!\hat{\sigma}_{j}^{x}(\hat{d}\!+\!\hat{d}^{\dagger})\!+\!\frac{\epsilon}{2}\!\sum_{j=1}^{N}\!\hat{\sigma}_{j}^{z}\!-\!\frac{J}{N}\!\sum_{j<k}\!\hat{\sigma}_{j}^{y}\hat{\sigma}_{k}^{y},\!\label{eq:H_full}
\end{equation} 
Here, $\hat{d}(\hat{d}^{\dagger})$ denotes the output bosonic mode. Its frequency has been taken as the unit of energy $\omega_{0}=1$ and all the other parameters in the Hamiltonian have been rescaled by $\omega_{0}$. The operators $\hat{\sigma}_{j}^{\alpha}\ (\alpha=x,y,z)$
are the Pauli matrices of the $j$th spin. A magnetic field is applied
along the $z$-direction inducing a energy splitting $\epsilon$ between
spin states $\left|\uparrow\right\rangle _{j}$ and $\left|\downarrow\right\rangle _{j}$. The spin-boson coupling is along $x$-direction with homogeneous coupling strength $\lambda$ similar to the Dicke model~\cite{dicke1954coherence}. The last term describes the all-to-all homogeneous coupling between the spins along the $y$-direction akin to the Lipkin-Meshkov-Glick (LMG)  model~\cite{lipkin1965validity,meshkov1965validity,glick1965validity}. Here, we call this model as Dicke-LMGy model and emphasize the critical distinction from the Dicke-Ising model which only has nearest neighbour spin interactions~\cite{lee2004first}.  The amplification process in the QCD is triggered by the weak input signal induced variation in the spin-boson coupling $\lambda$ or equivalently the spin-spin coupling $J$. 

\begin{figure}
\includegraphics[width=8.5cm]{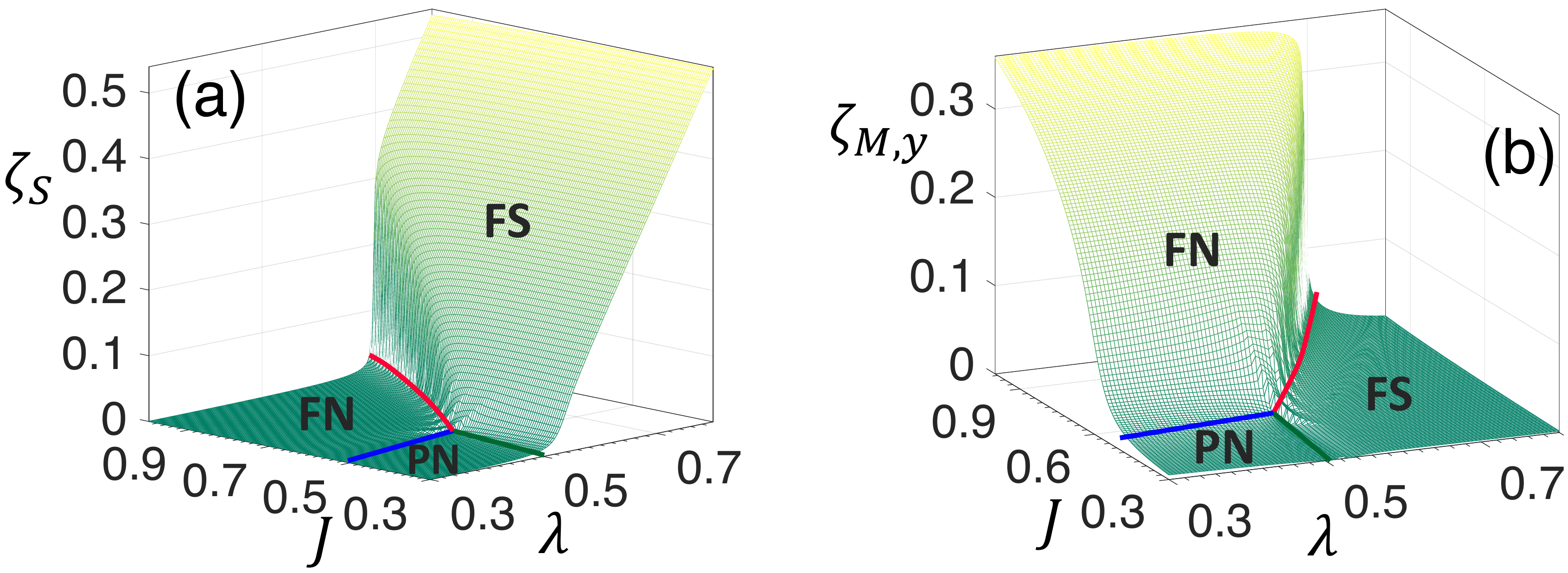}
\caption{\label{fig:NumericalPhaseTransition} Numerical demonstration of the
phase diagram with the superradiant order parameter $\zeta_S=\langle\hat{d}^{\dagger}\hat{d}\rangle_0/N$ in panel (a) and the magnetic order parameter $\zeta_{M,y}=\langle\hat{S}_{y}^{2}\rangle_0/N^{2}$ in panel (b). The boundaries between different phases are marked out by the blue, the green, and the red lines, which correspond to the three lines in Fig.~\ref{fig:schematic} (b) exactly.  In this figure, the other parameters are taken as $\epsilon =1$, and both the spin number $N$ and the cutoff for the bosonic mode are set as $40$.}
\end{figure}

To characterize the quantum phases and the corresponding QPTs in our Dicke-LMGy model, we introduce two new magnetic order parameters (OPs): $\zeta_{M,x}=\langle\hat{S}^{2}_x\rangle_0/N^2$ and $\zeta_{M,y}=\langle\hat{S}^{2}_y\rangle_0/N^2$ ($\hat{S}_{\alpha}=\sum_j \hat{\sigma}_j^{\alpha}/2$ and $\langle\cdots\rangle_0$ means averaging on the ground state) characterizing the magnetic fluctuations in the spins along $x$ and $y$ axes, respectively. Such OPs can be probed experimentally through spin noise spectroscopy~\cite{zapasskii2013spin}. Note, we do not choose the traditional magnetic OP $M_z=\langle\hat{S}_{z}\rangle_0/N$~\cite{pfeuty1970one}, due to its incapability of characterizing the first-order QPT in our model~\cite{yang2018numerical}. The superradiant OP $\zeta_S=\langle\hat{d}^{\dagger}\hat{d}\rangle_0/N$ is utilized to characterize the macroscopic excitation in the bosonic mode and functions as the output observable of our QCD.  There exist three quantum phases in our model: paramagnetic-normal (PN) phase, ferromagnetic-normal
(FN) phase, and ferromagnetic-superradiant (FS) phase as shown in Fig.~\ref{fig:schematic} (b). In the $\lambda J$-plane, the the blue line determined by the critical spin-spin coupling strength $J_{c,{\rm II}}\equiv\epsilon/2$, the green line determined by the critical spin-boson coupling $\lambda_{c,{\rm II}}\equiv\sqrt{\epsilon}/2$, and the red line $J=2\lambda^{2}$ give the phase boundaries. We emphasize that the FN-FS boundary displays a first order QPT which is necessary for our QCD. There exists a unique triple-point $(\lambda_{c,{\rm II}},\ J_{c,{\rm II}})$ determined by the energy splitting $\epsilon$. The numerical demonstration of the phase diagram is given in Fig.~\ref{fig:NumericalPhaseTransition} with the superradiant OP $\zeta_S$ and the magnetic OP $\zeta_{M,y}$ in panels (a) and (b), respectively. The OP $\zeta_{M,x}$ not shown here behaves similarly to $\zeta_S$. We exploit mean field theory ~\cite{yang2018numerical,zhang2014quantum} to verify this phase diagram obtained by our direct numerical calculation. We emphasize that the numerical approach we have introduced holds a fundamental advantage for dynamical amplification and noise calculations.

\begin{figure}
\includegraphics[width=8.5cm]{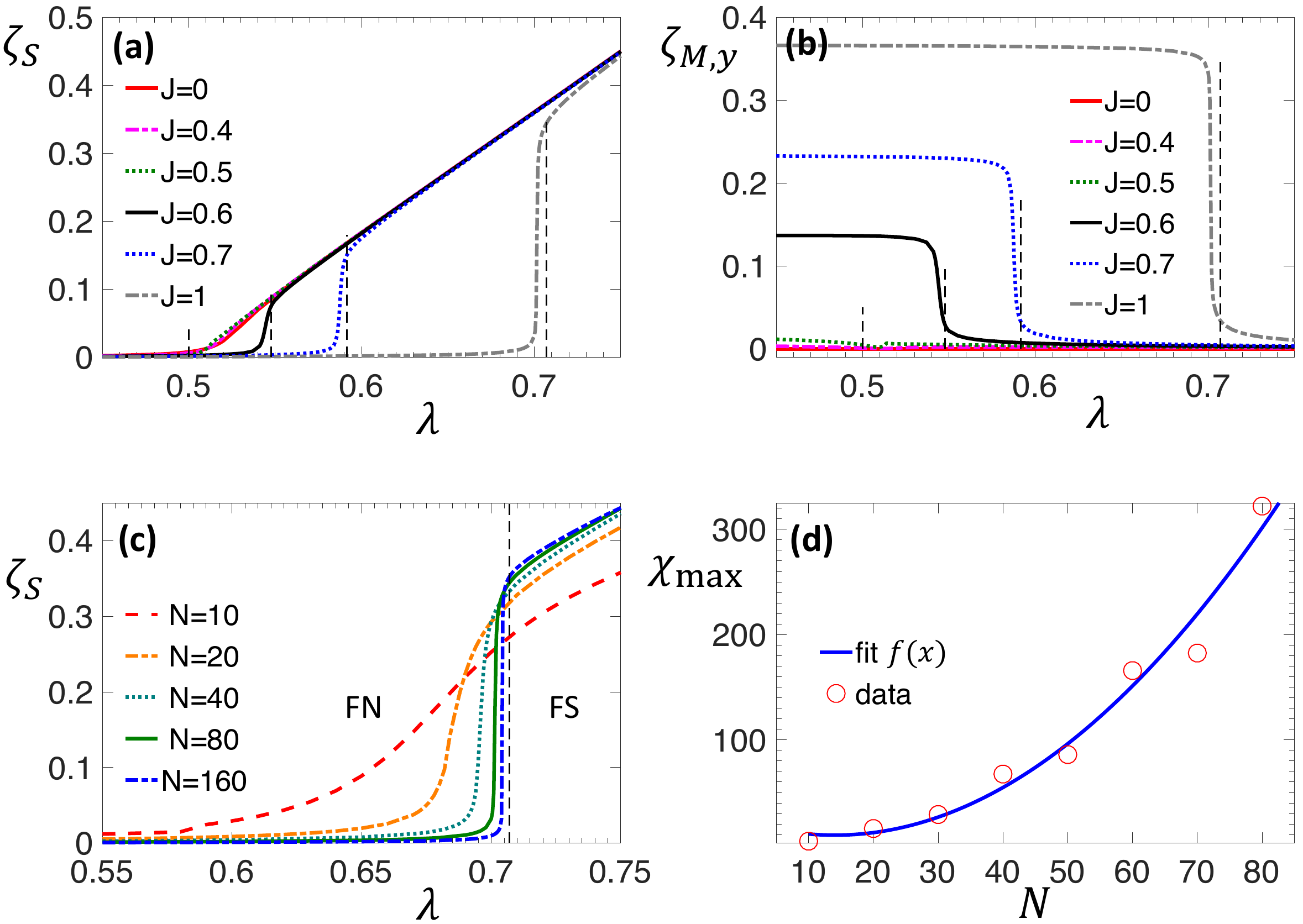}
\caption{\label{fig:SupRadPT} In (a) and (b), we show the order parameters $\zeta_S$ and $\zeta_{M,y}$, respectively, for different spin-spin coupling $J$. Here, $\epsilon=1$ and both the spin number and the cutoff of the bosonic mode are set as $80$. Second-order QPTs occur at $\lambda_{c,{\rm II}}\equiv\sqrt{\epsilon}/2=0.5$ for $J\leq J_{c,{\rm II}}\equiv\epsilon/2=0.5$ . While first-order QPTs occur at $\lambda_{c,{\rm I}}\equiv\sqrt{J/2}$ for $J>J_{c,{\rm II}}$. The locations of the critical points are marked by the thin black dashed line. In panel (c), we plot  $\zeta_S$ with fixed $J=1>J_{c,{\rm II}}$ for different spin number $N$. In panel (d), we show that the maximum of the sensitivity diverges with the spin number $N$ verifying the first-order QPT. The polynomial fitting function $f(x)=0.067x^{2}-1.881x+22.62$
shows the $N^{2}$ scaling.}
\end{figure}

The most striking property of our tractable model is that the QPT between the FN phase and the FS phase is of first order making QCD a highly sensitive device. This first order transition
occurs in the strong coupling regime for large spin-spin coupling  ($J>J_{c,{\rm II}}$) and spin-boson coupling ($\lambda > \lambda_{c,{\rm II}}$). As shown by the black, blue, and gray lines in Fig.~\ref{fig:SupRadPT} (a) and (b), discontinuous changes are observed in both the superradiant OP $\zeta_S$ and the magnetic OP $\zeta_{M,y}$. Note that the overlapping red, pink and green curves correspond to low spin-spin coupling giving rise only to a second order transition between the PN and FS phase, not suitable for QCD. Once the spin-spin coupling increases, there is no paramagnetic phase and there exists only a ferromagnetic phase for all spin-boson coupling strengths. This ferromagnetism is evident by studying the magnetic order parameter in  Fig.~\ref{fig:SupRadPT} (b) (black, blue and gray curves). We emphasize that the ferromagnetic behavior shown in Fig.~\ref{fig:SupRadPT} (b) is only along the y direction and the exact opposite trend occurs for the x direction (not shown). During the FN to FS transition which is first order, the energy prestored in the spins transfers to the bosonic mode coupled with a change in the spin fluctuations from the $y$-direction to the $x$-direction.  The first-order QPT results from the competition between the FS phase -- that arises from strong spin-boson coupling along the $x$-axis and the FN phase -- caused by large spin-spin coupling along the $y$-axis. We also predict that this type of first-order QPT should also exist in the Ising XY-model~\cite{Ovchinnikov2003anti-ferro}.  Another important characteristic is that the first-order phase transition point is sensitive to the spin-spin coupling (bias) while the second order ones in Fig.~\ref{fig:NumericalPhaseTransition} (a) is not.

We now explicitly show that the FN-FS phase transition is of first order in Fig.~\ref{fig:SupRadPT} (c) as required for QCD. Increasing the spin number, the phase transition shows critical scaling behavior. Here, the sensitivity of the system at the critical point is determined by first-order derivative of the superradiant OP $\zeta_S$,
\begin{equation}
\chi(\lambda)=\frac{1}{N}\frac{d}{d\lambda}\langle\hat{d}^{\dagger}\hat{d}\rangle,
\end{equation}
where the factor $1/N$ is for consistency with the magnetic susceptibility. In Fig.~\ref{fig:SupRadPT} (d), we plot the maximum of the sensitivity function ($\chi_{{\rm max}}$) at the first-order critical point $\lambda_{c,{\rm I}}\equiv \sqrt{J/2}$ vs spin number ($N$). In the thermodynamic limit $N\rightarrow\infty$, $\chi_{{\rm max}}$ diverges with speed $\propto N^{2}$, which
is much faster than the $\sqrt{N}$-scaling obtained in the previous
first-order dissipative transition~\cite{raghunandan2018high} or
the linear $N$ scaling in the first-order thermodynamic phase transition
predicted by Imry~\cite{imry1980finite}. This $N^{2}$ scaling, unique to our model arising from competing phases, may
be used to enhance the sensitivity~\cite{skotiniotis2015quantum} and beat the Heisenberg limit in parameter estimation~\cite{supplementary}. 

We emphasize that our proposed model  has  two  main  applications:  (1)  quantum  parameter  estimation (time-independent  process)  with $N^2$-scaling~\cite{supplementary};  (2) quantum  dynamical  amplification  (time-dependent  process)  with $N$-scaling as shown in the following.

\textit{Dynamical quantum amplification}---To utilize quantum criticality as a resource of quantum amplification, one has to study the dynamical behavior around the critical point. The challenge in exploiting QPTs for quantum amplification arises fundamentally from the fact that ground states of two different phases can not be connected via an adiabatic operation~\cite{dziarmaga2010dynamics}. For example, starting from the ground state of the FN phase, the system can alternatively evolve to an arbitrary excited state instead of going to the ground state of the FS phase thereby completely negating critical amplification. To demonstrate the dynamical nonlinear amplification, here we show the dynamics of the first-order QPT in our model with 80 spins via  direct numerical time-evolution. A QCD is demonstrated with the large quantum gain around the critical point after a DQPT. We also show that a linear scaling in the quantum gain and SQNR of the QCD is obtained, instead of the $N^2$ in the sensitivity of the first-order QPT. 

\begin{figure}
\includegraphics[width=7cm]{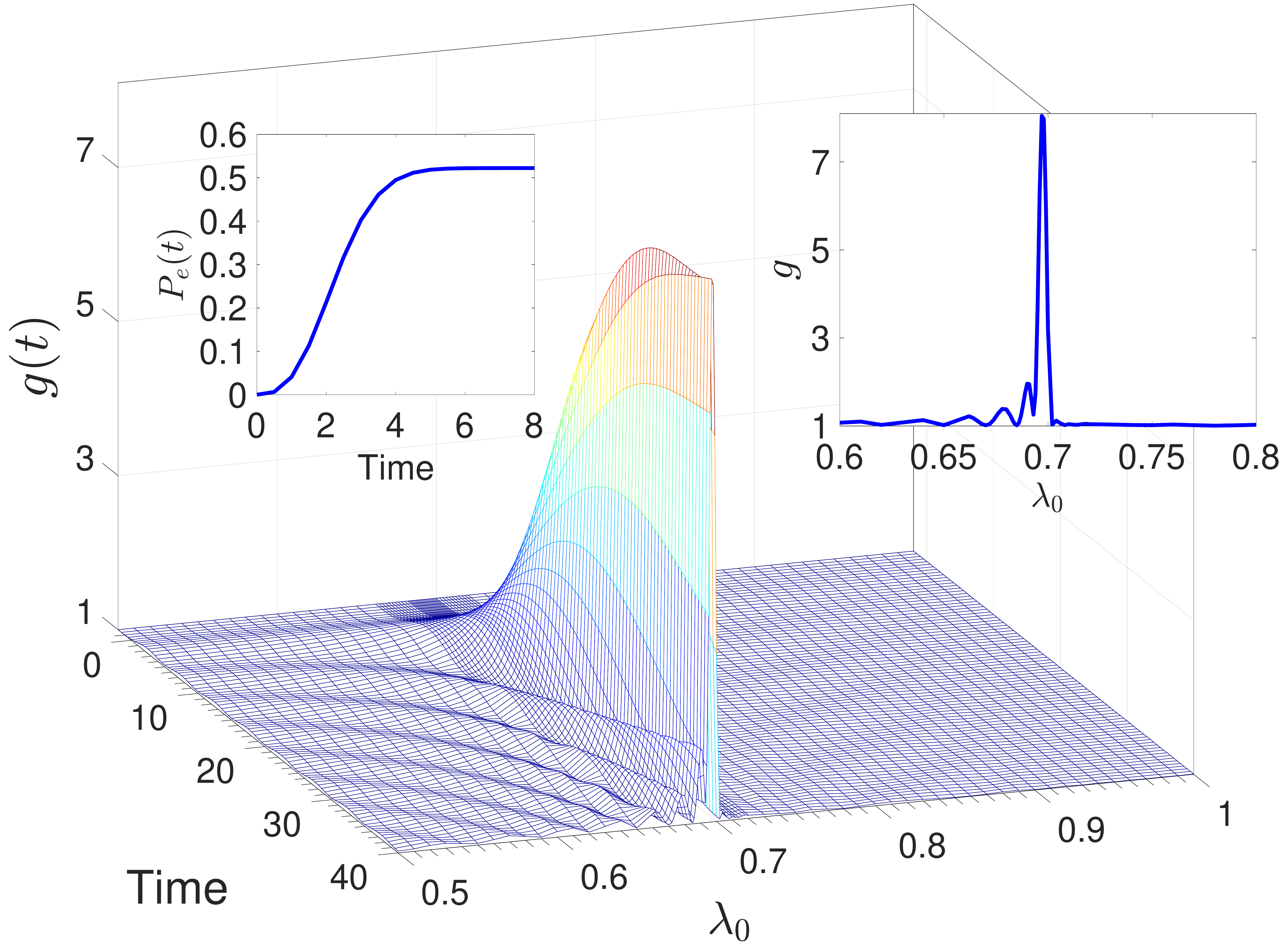}\caption{\label{fig:dynamic_QPT}The  amplification via the first-order dynamical quantum phase transition is shown by the time-dependent quantum gain $g(t)$. The left subplot shows envelope $P_{e}(t)$ in the time dependent spin-boson coupling $\lambda(t)$. Here, the spin-spin
coupling is set as $J=1>J_{c,{\rm II}}$, the time is in a unit of $1/\omega_{0}$,
and amplitude of the small change in the parameter $\lambda(t)$ is set
to be $\Delta\lambda=0.01$. In the right subplot, we show that the
quantum amplification only occurs when $\lambda_{0}$ is biased close
to critical point $\lambda_{c,{\rm I}}=\sqrt{J/2}\approx0.707$ with $t=40$. }
\end{figure}

The full measurement in our QCD is split into two main processes: transduction (absorption) and amplification. After the transduction, the excitations or energy in the input signal will be transferred into the detector inducing a time-dependent variation of the spin-boson coupling strength $\lambda(t)=\lambda_{0}+\Delta\lambda\times P_{e}(t)$. The key amplification process in our QCD occurs from first-order DQPT process triggered by this time-dependent parameter $\lambda (t)$. The detector is optimally biased so that the spin-boson coupling $\lambda_0$ lies very close to the critical point $\lambda_{c,{\rm I}}$. Thus, even a very small parameter variation $\Delta\lambda$ (amplitude) can trigger a DQPT and leads to efficient amplification. As mentioned before, amplification is defined as the enhanced number of bosons in the cavity mode. The time-dependent envelope $P_{e}(t)$ of the coupling $\lambda(t)$ is determined by the transduction process~\cite{supplementary}. One example of such a weak input signal is the single-photon absorption probability~\cite{yang2018concept} or the wave form of the excitation in a detector. The full dynamics of the whole system is governed by the time-dependent Hamiltonian $H(t)$ in Eq.~(\ref{eq:H_full}) by replacing $\lambda$ with $\lambda(t)$. To characterize the detection sensitivity, we define the quantum gain of the amplification  in our QCD as
\begin{equation}
g(t)=\langle\hat{d}^{\dagger}(t)\hat{d}(t)\rangle/\langle\hat{d}^{\dagger}(0)\hat{d}(0)\rangle,
\end{equation}
where $\langle\hat{d}^{\dagger}(t)\hat{d}(t)\rangle$ is the mean
value of the time-dependent operator in the Heisenberg picture on
the ground state of initial Hamiltonian $H(0)$ with $\lambda=\lambda_{0}$.

\begin{figure}
\includegraphics[width=7cm]{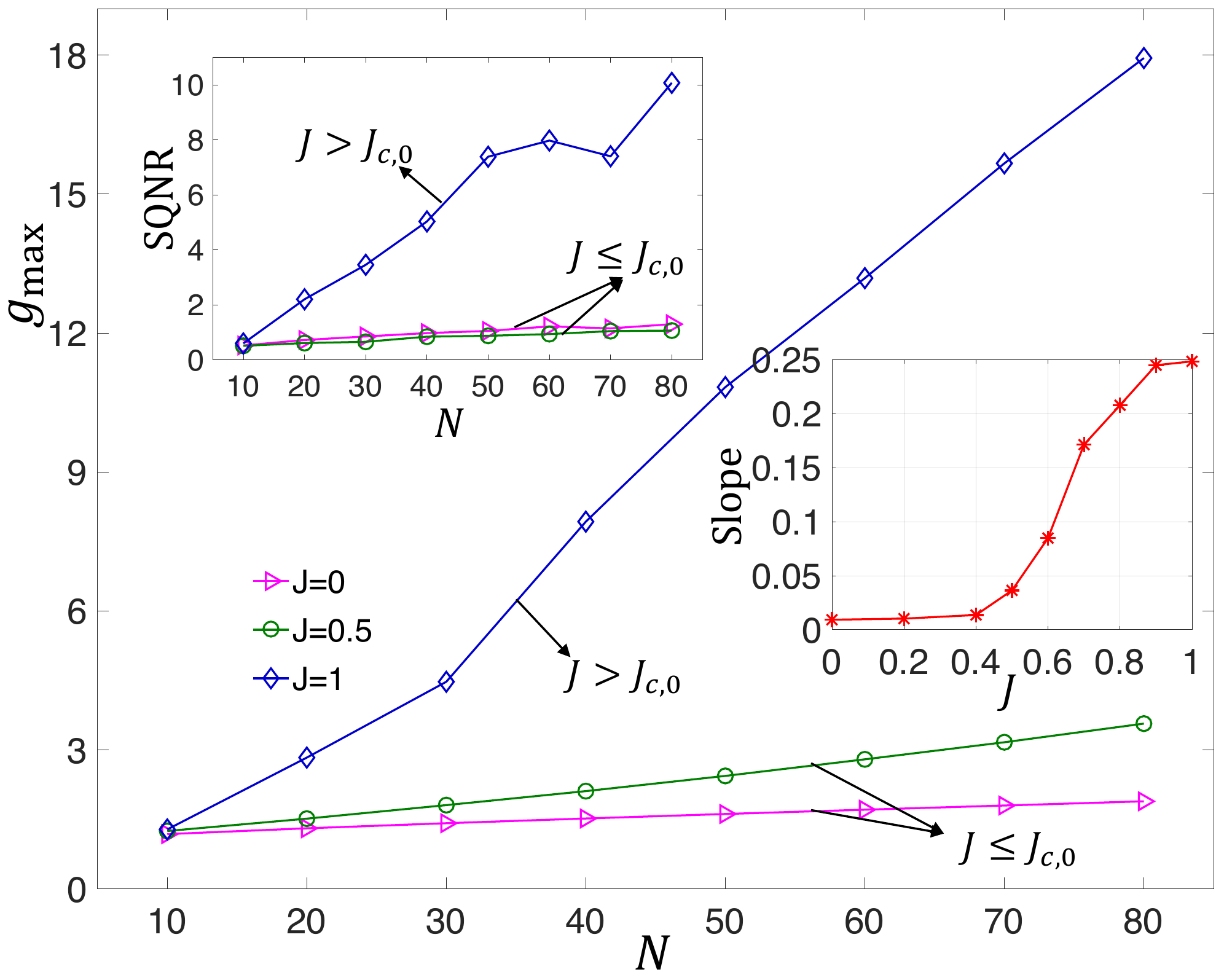}\caption{\label{fig:Scaling}Linear scaling of the quantum bosonic amplification
with the spin number $N$ for different spin-spin coupling $J$. The
corresponding signal-to-quantum noise ratios are shown in the left
subplot. In the right subplot, we shown there is a phase-transition-like
behavior in the slope of the maximum gain when $J$ crosses the critical
point $J_{c,{\rm II}}=0.5$. First-order dynamic QPT (blue diamond line) has
much higher gain and signal-to-quantum noise ratio than that of second-order
QPT.}
\end{figure}

The dynamical amplification in the QCD is demonstrated by the time-dependent gain as a function of the bias spin-boson coupling $\lambda_{0}$ and time in Fig.~\ref{fig:dynamic_QPT}. We see that the efficient amplification can only be obtained if the system is optimally biased close to the critical point. In the right subplot computed at fixed time $t=40$, we explicitly show the high gain of our QCD around the critical point. Here, we show the highly sensitive nature of a time-dependent first-order QPT system to the initial bias. Similar to the enhanced decay of the Loschmidt echo by the criticality in a second-order QPT~\cite{quan2006decay}, the enhanced quantum gain in our QCD is a universal characteristic of criticality in a first-order DQPT. We emphasize that this universal characteristic is fundamentally different from the ones discovered in previous second-order DQPTs~\cite{Dziarmaga2005Dynamics,Zurek2005Dynamics,Heyl2013Dynamical,Abeling2016quantum,Zunkovic2018dynamical,pelissetto2018out}, such as the Kibble-Zurek mechanism~\cite{dziarmaga2010dynamics} or the non-analytic kinks in the logarithm of Loschmidt echo~\cite{heyl2018dynamical}. 

To show the high figures of merit of our QCD, we present the scaling of the quantum amplification with the spin number $N$ in Fig.~\ref{fig:Scaling}. The maximum gain is shown to be linearly
proportional to $N$. This result should be contrasted with the amplification resulting from the second
order phase transition with $J\leq J_{c,{\rm II}}$ (see pink triangle line
and the green circle line). The latter is much smaller than the one from the first-order DQPT with $J>J_{c,{\rm II}}$ (see the blue diamond line). In the right
subplot, we show the slope of $g_{{\rm max}}$ with the spin number
for different spin-spin coupling. There exists a ``phase transition''
phenomenon in the slope at the same critical point $J_{c,{\rm II}}$ of
the transition from second-order to first-order QPTs. To characterize
the quantum noise in our QCD, we define the SQNR as~\cite{yuen1976states}
\begin{equation}
{\rm SQNR}=\langle\hat{d}^{\dagger}(t)\hat{d}(t)\rangle^{2}/\langle[\Delta\hat{d}^{\dagger}(t)\hat{d}(t)]^{2}\rangle,
\end{equation}
where $\langle[\Delta\hat{d}^{\dagger}(t)\hat{d}(t)]^{2}\rangle=\langle[\hat{d}^{\dagger}(t)\hat{d}(t)]^{2}\rangle-\langle\hat{d}^{\dagger}(t)\hat{d}(t)\rangle^{2}$
is the variance of the bosonic excitation number operator. The corresponding
SQNR for the three lines are displayed in the left subplot. It shows
amplification based on first-order QPT has much higher SQNR than that
of second-order QPTs. Similar to the quantum gain (the re-scaled excitation number in the output bosonic mode), the SQNR also increases linearly with the spin number, which is consistent with the SQNR of a coherent state as the final output state. 

\textit{Discussion}---In recent experiments, second-order DQPTs have been demonstrated with Bose-Einstein condensates ~\cite{baumann2010dicke}, trapped ions \cite{zhang2017observation,Jurcevic2017direct}, cold atoms \cite{bernien2017probing}, and superconducting qubits~\cite{Harris2018Phase}. On the other hand, the quantum critical detector requires engineered first order phase transitions. However, we believe these platforms provide a starting point for physical realization of our model where single photon  perturbations can trigger a phase transition in systems with long-range interactions. 

We thank C. L. Cortes and D.-Z. Xu for fruitful discussions. This work is supported by DARPA DETECT. 

\bibliography{main}

\begin{thebibliography}{56}%
\makeatletter
\providecommand \@ifxundefined [1]{%
 \@ifx{#1\undefined}
}%
\providecommand \@ifnum [1]{%
 \ifnum #1\expandafter \@firstoftwo
 \else \expandafter \@secondoftwo
 \fi
}%
\providecommand \@ifx [1]{%
 \ifx #1\expandafter \@firstoftwo
 \else \expandafter \@secondoftwo
 \fi
}%
\providecommand \natexlab [1]{#1}%
\providecommand \enquote  [1]{``#1''}%
\providecommand \bibnamefont  [1]{#1}%
\providecommand \bibfnamefont [1]{#1}%
\providecommand \citenamefont [1]{#1}%
\providecommand \href@noop [0]{\@secondoftwo}%
\providecommand \href [0]{\begingroup \@sanitize@url \@href}%
\providecommand \@href[1]{\@@startlink{#1}\@@href}%
\providecommand \@@href[1]{\endgroup#1\@@endlink}%
\providecommand \@sanitize@url [0]{\catcode `\\12\catcode `\$12\catcode
  `\&12\catcode `\#12\catcode `\^12\catcode `\_12\catcode `\%12\relax}%
\providecommand \@@startlink[1]{}%
\providecommand \@@endlink[0]{}%
\providecommand \url  [0]{\begingroup\@sanitize@url \@url }%
\providecommand \@url [1]{\endgroup\@href {#1}{\urlprefix }}%
\providecommand \urlprefix  [0]{URL }%
\providecommand \Eprint [0]{\href }%
\providecommand \doibase [0]{http://dx.doi.org/}%
\providecommand \selectlanguage [0]{\@gobble}%
\providecommand \bibinfo  [0]{\@secondoftwo}%
\providecommand \bibfield  [0]{\@secondoftwo}%
\providecommand \translation [1]{[#1]}%
\providecommand \BibitemOpen [0]{}%
\providecommand \bibitemStop [0]{}%
\providecommand \bibitemNoStop [0]{.\EOS\space}%
\providecommand \EOS [0]{\spacefactor3000\relax}%
\providecommand \BibitemShut  [1]{\csname bibitem#1\endcsname}%
\let\auto@bib@innerbib\@empty
\bibitem [{\citenamefont {Caves}(1982)}]{caves1982quantum}%
  \BibitemOpen
  \bibfield  {author} {\bibinfo {author} {\bibfnamefont {C.~M.}\ \bibnamefont
  {Caves}},\ }\href
  {https://journals.aps.org/prd/abstract/10.1103/PhysRevD.26.1817} {\bibfield
  {journal} {\bibinfo  {journal} {Physical Review D}\ }\textbf {\bibinfo
  {volume} {26}},\ \bibinfo {pages} {1817} (\bibinfo {year}
  {1982})}\BibitemShut {NoStop}%
\bibitem [{\citenamefont {Louisell}\ \emph {et~al.}(1961)\citenamefont
  {Louisell}, \citenamefont {Yariv},\ and\ \citenamefont
  {Siegman}}]{louisell1961quantum}%
  \BibitemOpen
  \bibfield  {author} {\bibinfo {author} {\bibfnamefont {W.~H.}\ \bibnamefont
  {Louisell}}, \bibinfo {author} {\bibfnamefont {A.}~\bibnamefont {Yariv}}, \
  and\ \bibinfo {author} {\bibfnamefont {A.~E.}\ \bibnamefont {Siegman}},\
  }\href {\doibase 10.1103/PhysRev.124.1646} {\bibfield  {journal} {\bibinfo
  {journal} {Phys. Rev.}\ }\textbf {\bibinfo {volume} {124}},\ \bibinfo {pages}
  {1646} (\bibinfo {year} {1961})}\BibitemShut {NoStop}%
\bibitem [{\citenamefont {Mollow}\ and\ \citenamefont
  {Glauber}(1967)}]{Mollow1967parametric1}%
  \BibitemOpen
  \bibfield  {author} {\bibinfo {author} {\bibfnamefont {B.~R.}\ \bibnamefont
  {Mollow}}\ and\ \bibinfo {author} {\bibfnamefont {R.~J.}\ \bibnamefont
  {Glauber}},\ }\href {\doibase 10.1103/PhysRev.160.1076} {\bibfield  {journal}
  {\bibinfo  {journal} {Phys. Rev.}\ }\textbf {\bibinfo {volume} {160}},\
  \bibinfo {pages} {1076} (\bibinfo {year} {1967})}\BibitemShut {NoStop}%
\bibitem [{\citenamefont {Gavish}\ \emph {et~al.}(2004)\citenamefont {Gavish},
  \citenamefont {Yurke},\ and\ \citenamefont {Imry}}]{Gavish2004generalized}%
  \BibitemOpen
  \bibfield  {author} {\bibinfo {author} {\bibfnamefont {U.}~\bibnamefont
  {Gavish}}, \bibinfo {author} {\bibfnamefont {B.}~\bibnamefont {Yurke}}, \
  and\ \bibinfo {author} {\bibfnamefont {Y.}~\bibnamefont {Imry}},\ }\href
  {\doibase 10.1103/PhysRevLett.93.250601} {\bibfield  {journal} {\bibinfo
  {journal} {Phys. Rev. Lett.}\ }\textbf {\bibinfo {volume} {93}},\ \bibinfo
  {pages} {250601} (\bibinfo {year} {2004})}\BibitemShut {NoStop}%
\bibitem [{\citenamefont {Roy}\ and\ \citenamefont
  {Devoret}(2016)}]{roy2016introduction}%
  \BibitemOpen
  \bibfield  {author} {\bibinfo {author} {\bibfnamefont {A.}~\bibnamefont
  {Roy}}\ and\ \bibinfo {author} {\bibfnamefont {M.}~\bibnamefont {Devoret}},\
  }\href {https://www.sciencedirect.com/science/article/pii/S1631070516300640}
  {\bibfield  {journal} {\bibinfo  {journal} {Comptes Rendus Physique}\
  }\textbf {\bibinfo {volume} {17}},\ \bibinfo {pages} {740} (\bibinfo {year}
  {2016})}\BibitemShut {NoStop}%
\bibitem [{\citenamefont {Devoret}\ and\ \citenamefont
  {Schoelkopf}(2000)}]{devoret2000amplifying}%
  \BibitemOpen
  \bibfield  {author} {\bibinfo {author} {\bibfnamefont {M.~H.}\ \bibnamefont
  {Devoret}}\ and\ \bibinfo {author} {\bibfnamefont {R.~J.}\ \bibnamefont
  {Schoelkopf}},\ }\href {https://www.nature.com/articles/35023253} {\bibfield
  {journal} {\bibinfo  {journal} {Nature}\ }\textbf {\bibinfo {volume} {406}},\
  \bibinfo {pages} {1039} (\bibinfo {year} {2000})}\BibitemShut {NoStop}%
\bibitem [{\citenamefont {Eisaman}\ \emph {et~al.}(2011)\citenamefont
  {Eisaman}, \citenamefont {Fan}, \citenamefont {Migdall},\ and\ \citenamefont
  {Polyakov}}]{eisaman2011invited}%
  \BibitemOpen
  \bibfield  {author} {\bibinfo {author} {\bibfnamefont {M.}~\bibnamefont
  {Eisaman}}, \bibinfo {author} {\bibfnamefont {J.}~\bibnamefont {Fan}},
  \bibinfo {author} {\bibfnamefont {A.}~\bibnamefont {Migdall}}, \ and\
  \bibinfo {author} {\bibfnamefont {S.~V.}\ \bibnamefont {Polyakov}},\ }\href
  {https://aip.scitation.org/doi/abs/10.1063/1.3610677} {\bibfield  {journal}
  {\bibinfo  {journal} {Review of scientific instruments}\ }\textbf {\bibinfo
  {volume} {82}},\ \bibinfo {pages} {071101} (\bibinfo {year}
  {2011})}\BibitemShut {NoStop}%
\bibitem [{\citenamefont {Gol’Tsman}\ \emph {et~al.}(2001)\citenamefont
  {Gol’Tsman}, \citenamefont {Okunev}, \citenamefont {Chulkova},
  \citenamefont {Lipatov}, \citenamefont {Semenov}, \citenamefont {Smirnov},
  \citenamefont {Voronov}, \citenamefont {Dzardanov}, \citenamefont
  {Williams},\ and\ \citenamefont {Sobolewski}}]{gol2001picosecond}%
  \BibitemOpen
  \bibfield  {author} {\bibinfo {author} {\bibfnamefont {G.}~\bibnamefont
  {Gol’Tsman}}, \bibinfo {author} {\bibfnamefont {O.}~\bibnamefont {Okunev}},
  \bibinfo {author} {\bibfnamefont {G.}~\bibnamefont {Chulkova}}, \bibinfo
  {author} {\bibfnamefont {A.}~\bibnamefont {Lipatov}}, \bibinfo {author}
  {\bibfnamefont {A.}~\bibnamefont {Semenov}}, \bibinfo {author} {\bibfnamefont
  {K.}~\bibnamefont {Smirnov}}, \bibinfo {author} {\bibfnamefont
  {B.}~\bibnamefont {Voronov}}, \bibinfo {author} {\bibfnamefont
  {A.}~\bibnamefont {Dzardanov}}, \bibinfo {author} {\bibfnamefont
  {C.}~\bibnamefont {Williams}}, \ and\ \bibinfo {author} {\bibfnamefont
  {R.}~\bibnamefont {Sobolewski}},\ }\href
  {https://aip.scitation.org/doi/abs/10.1063/1.1388868} {\bibfield  {journal}
  {\bibinfo  {journal} {Applied physics letters}\ }\textbf {\bibinfo {volume}
  {79}},\ \bibinfo {pages} {705} (\bibinfo {year} {2001})}\BibitemShut
  {NoStop}%
\bibitem [{\citenamefont {Sachdev}(2007)}]{sachdev2007quantum}%
  \BibitemOpen
  \bibfield  {author} {\bibinfo {author} {\bibfnamefont {S.}~\bibnamefont
  {Sachdev}},\ }\href
  {https://onlinelibrary.wiley.com/doi/full/10.1002/9780470022184.hmm108}
  {\emph {\bibinfo {title} {Quantum phase transitions}}}\ (\bibinfo
  {publisher} {Wiley Online Library},\ \bibinfo {year} {2007})\BibitemShut
  {NoStop}%
\bibitem [{\citenamefont {Quan}\ \emph {et~al.}(2006)\citenamefont {Quan},
  \citenamefont {Song}, \citenamefont {Liu}, \citenamefont {Zanardi},\ and\
  \citenamefont {Sun}}]{quan2006decay}%
  \BibitemOpen
  \bibfield  {author} {\bibinfo {author} {\bibfnamefont {H.~T.}\ \bibnamefont
  {Quan}}, \bibinfo {author} {\bibfnamefont {Z.}~\bibnamefont {Song}}, \bibinfo
  {author} {\bibfnamefont {X.~F.}\ \bibnamefont {Liu}}, \bibinfo {author}
  {\bibfnamefont {P.}~\bibnamefont {Zanardi}}, \ and\ \bibinfo {author}
  {\bibfnamefont {C.~P.}\ \bibnamefont {Sun}},\ }\href {\doibase
  10.1103/PhysRevLett.96.140604} {\bibfield  {journal} {\bibinfo  {journal}
  {Phys. Rev. Lett.}\ }\textbf {\bibinfo {volume} {96}},\ \bibinfo {pages}
  {140604} (\bibinfo {year} {2006})}\BibitemShut {NoStop}%
\bibitem [{\citenamefont {Lieb}\ \emph {et~al.}(1961)\citenamefont {Lieb},
  \citenamefont {Schultz},\ and\ \citenamefont {Mattis}}]{lieb1961two}%
  \BibitemOpen
  \bibfield  {author} {\bibinfo {author} {\bibfnamefont {E.}~\bibnamefont
  {Lieb}}, \bibinfo {author} {\bibfnamefont {T.}~\bibnamefont {Schultz}}, \
  and\ \bibinfo {author} {\bibfnamefont {D.}~\bibnamefont {Mattis}},\ }\href
  {https://www.sciencedirect.com/science/article/pii/0003491661901154}
  {\bibfield  {journal} {\bibinfo  {journal} {Annals of Physics}\ }\textbf
  {\bibinfo {volume} {16}},\ \bibinfo {pages} {407} (\bibinfo {year}
  {1961})}\BibitemShut {NoStop}%
\bibitem [{\citenamefont {Lipkin}\ \emph {et~al.}(1965)\citenamefont {Lipkin},
  \citenamefont {Meshkov},\ and\ \citenamefont {Glick}}]{lipkin1965validity}%
  \BibitemOpen
  \bibfield  {author} {\bibinfo {author} {\bibfnamefont {H.~J.}\ \bibnamefont
  {Lipkin}}, \bibinfo {author} {\bibfnamefont {N.}~\bibnamefont {Meshkov}}, \
  and\ \bibinfo {author} {\bibfnamefont {A.}~\bibnamefont {Glick}},\ }\href
  {https://www.sciencedirect.com/science/article/pii/002955826590862X}
  {\bibfield  {journal} {\bibinfo  {journal} {Nuclear Physics}\ }\textbf
  {\bibinfo {volume} {62}},\ \bibinfo {pages} {188} (\bibinfo {year}
  {1965})}\BibitemShut {NoStop}%
\bibitem [{\citenamefont {Meshkov}\ \emph {et~al.}(1965)\citenamefont
  {Meshkov}, \citenamefont {Glick},\ and\ \citenamefont
  {Lipkin}}]{meshkov1965validity}%
  \BibitemOpen
  \bibfield  {author} {\bibinfo {author} {\bibfnamefont {N.}~\bibnamefont
  {Meshkov}}, \bibinfo {author} {\bibfnamefont {A.}~\bibnamefont {Glick}}, \
  and\ \bibinfo {author} {\bibfnamefont {H.}~\bibnamefont {Lipkin}},\ }\href
  {https://www.sciencedirect.com/science/article/pii/0029558265908631}
  {\bibfield  {journal} {\bibinfo  {journal} {Nuclear Physics}\ }\textbf
  {\bibinfo {volume} {62}},\ \bibinfo {pages} {199} (\bibinfo {year}
  {1965})}\BibitemShut {NoStop}%
\bibitem [{\citenamefont {Glick}\ \emph {et~al.}(1965)\citenamefont {Glick},
  \citenamefont {Lipkin},\ and\ \citenamefont {Meshkov}}]{glick1965validity}%
  \BibitemOpen
  \bibfield  {author} {\bibinfo {author} {\bibfnamefont {A.}~\bibnamefont
  {Glick}}, \bibinfo {author} {\bibfnamefont {H.}~\bibnamefont {Lipkin}}, \
  and\ \bibinfo {author} {\bibfnamefont {N.}~\bibnamefont {Meshkov}},\ }\href
  {https://www.sciencedirect.com/science/article/pii/0029558265908643}
  {\bibfield  {journal} {\bibinfo  {journal} {Nuclear Physics}\ }\textbf
  {\bibinfo {volume} {62}},\ \bibinfo {pages} {211} (\bibinfo {year}
  {1965})}\BibitemShut {NoStop}%
\bibitem [{\citenamefont {Hepp}\ and\ \citenamefont
  {Lieb}(1973)}]{hepp1973superradiant}%
  \BibitemOpen
  \bibfield  {author} {\bibinfo {author} {\bibfnamefont {K.}~\bibnamefont
  {Hepp}}\ and\ \bibinfo {author} {\bibfnamefont {E.~H.}\ \bibnamefont
  {Lieb}},\ }\href
  {https://www.sciencedirect.com/science/article/pii/0003491673900390}
  {\bibfield  {journal} {\bibinfo  {journal} {Annals of Physics}\ }\textbf
  {\bibinfo {volume} {76}},\ \bibinfo {pages} {360} (\bibinfo {year}
  {1973})}\BibitemShut {NoStop}%
\bibitem [{\citenamefont {Wang}\ and\ \citenamefont
  {Hioe}(1973)}]{wang1973phase}%
  \BibitemOpen
  \bibfield  {author} {\bibinfo {author} {\bibfnamefont {Y.~K.}\ \bibnamefont
  {Wang}}\ and\ \bibinfo {author} {\bibfnamefont {F.}~\bibnamefont {Hioe}},\
  }\href {https://journals.aps.org/pra/abstract/10.1103/PhysRevA.7.831}
  {\bibfield  {journal} {\bibinfo  {journal} {Physical Review A}\ }\textbf
  {\bibinfo {volume} {7}},\ \bibinfo {pages} {831} (\bibinfo {year}
  {1973})}\BibitemShut {NoStop}%
\bibitem [{\citenamefont {Giovannetti}\ \emph {et~al.}(2004)\citenamefont
  {Giovannetti}, \citenamefont {Lloyd},\ and\ \citenamefont
  {Maccone}}]{giovannetti2004quantum}%
  \BibitemOpen
  \bibfield  {author} {\bibinfo {author} {\bibfnamefont {V.}~\bibnamefont
  {Giovannetti}}, \bibinfo {author} {\bibfnamefont {S.}~\bibnamefont {Lloyd}},
  \ and\ \bibinfo {author} {\bibfnamefont {L.}~\bibnamefont {Maccone}},\ }\href
  {http://science.sciencemag.org/content/306/5700/1330} {\bibfield  {journal}
  {\bibinfo  {journal} {Science}\ }\textbf {\bibinfo {volume} {306}},\ \bibinfo
  {pages} {1330} (\bibinfo {year} {2004})}\BibitemShut {NoStop}%
\bibitem [{\citenamefont {Giovannetti}\ \emph {et~al.}(2006)\citenamefont
  {Giovannetti}, \citenamefont {Lloyd},\ and\ \citenamefont
  {Maccone}}]{giovannetti2006quantum}%
  \BibitemOpen
  \bibfield  {author} {\bibinfo {author} {\bibfnamefont {V.}~\bibnamefont
  {Giovannetti}}, \bibinfo {author} {\bibfnamefont {S.}~\bibnamefont {Lloyd}},
  \ and\ \bibinfo {author} {\bibfnamefont {L.}~\bibnamefont {Maccone}},\ }\href
  {https://journals.aps.org/prl/abstract/10.1103/PhysRevLett.96.010401}
  {\bibfield  {journal} {\bibinfo  {journal} {Physical review letters}\
  }\textbf {\bibinfo {volume} {96}},\ \bibinfo {pages} {010401} (\bibinfo
  {year} {2006})}\BibitemShut {NoStop}%
\bibitem [{\citenamefont {Lee}\ and\ \citenamefont
  {Johnson}(2004)}]{lee2004first}%
  \BibitemOpen
  \bibfield  {author} {\bibinfo {author} {\bibfnamefont {C.~F.}\ \bibnamefont
  {Lee}}\ and\ \bibinfo {author} {\bibfnamefont {N.~F.}\ \bibnamefont
  {Johnson}},\ }\href
  {https://journals.aps.org/prl/abstract/10.1103/PhysRevLett.93.083001}
  {\bibfield  {journal} {\bibinfo  {journal} {Physical review letters}\
  }\textbf {\bibinfo {volume} {93}},\ \bibinfo {pages} {083001} (\bibinfo
  {year} {2004})}\BibitemShut {NoStop}%
\bibitem [{\citenamefont {Ovchinnikov}\ \emph {et~al.}(2003)\citenamefont
  {Ovchinnikov}, \citenamefont {Dmitriev}, \citenamefont {Krivnov},\ and\
  \citenamefont {Cheranovskii}}]{Ovchinnikov2003anti-ferro}%
  \BibitemOpen
  \bibfield  {author} {\bibinfo {author} {\bibfnamefont {A.~A.}\ \bibnamefont
  {Ovchinnikov}}, \bibinfo {author} {\bibfnamefont {D.~V.}\ \bibnamefont
  {Dmitriev}}, \bibinfo {author} {\bibfnamefont {V.~Y.}\ \bibnamefont
  {Krivnov}}, \ and\ \bibinfo {author} {\bibfnamefont {V.~O.}\ \bibnamefont
  {Cheranovskii}},\ }\href {\doibase 10.1103/PhysRevB.68.214406} {\bibfield
  {journal} {\bibinfo  {journal} {Phys. Rev. B}\ }\textbf {\bibinfo {volume}
  {68}},\ \bibinfo {pages} {214406} (\bibinfo {year} {2003})}\BibitemShut
  {NoStop}%
\bibitem [{\citenamefont {Vidal}\ \emph {et~al.}(2004)\citenamefont {Vidal},
  \citenamefont {Mosseri},\ and\ \citenamefont
  {Dukelsky}}]{vidal2004entanglement}%
  \BibitemOpen
  \bibfield  {author} {\bibinfo {author} {\bibfnamefont {J.}~\bibnamefont
  {Vidal}}, \bibinfo {author} {\bibfnamefont {R.}~\bibnamefont {Mosseri}}, \
  and\ \bibinfo {author} {\bibfnamefont {J.}~\bibnamefont {Dukelsky}},\ }\href
  {https://journals.aps.org/pra/abstract/10.1103/PhysRevA.69.054101} {\bibfield
   {journal} {\bibinfo  {journal} {Physical Review A}\ }\textbf {\bibinfo
  {volume} {69}},\ \bibinfo {pages} {054101} (\bibinfo {year}
  {2004})}\BibitemShut {NoStop}%
\bibitem [{\citenamefont {Del~Re}\ \emph {et~al.}(2016)\citenamefont {Del~Re},
  \citenamefont {Fabrizio},\ and\ \citenamefont
  {Tosatti}}]{del2016nonequilium}%
  \BibitemOpen
  \bibfield  {author} {\bibinfo {author} {\bibfnamefont {L.}~\bibnamefont
  {Del~Re}}, \bibinfo {author} {\bibfnamefont {M.}~\bibnamefont {Fabrizio}}, \
  and\ \bibinfo {author} {\bibfnamefont {E.}~\bibnamefont {Tosatti}},\ }\href
  {\doibase 10.1103/PhysRevB.93.125131} {\bibfield  {journal} {\bibinfo
  {journal} {Phys. Rev. B}\ }\textbf {\bibinfo {volume} {93}},\ \bibinfo
  {pages} {125131} (\bibinfo {year} {2016})}\BibitemShut {NoStop}%
\bibitem [{\citenamefont {Gammelmark}\ and\ \citenamefont
  {M{\o}lmer}(2011)}]{gammelmark2011phase}%
  \BibitemOpen
  \bibfield  {author} {\bibinfo {author} {\bibfnamefont {S.}~\bibnamefont
  {Gammelmark}}\ and\ \bibinfo {author} {\bibfnamefont {K.}~\bibnamefont
  {M{\o}lmer}},\ }\href
  {http://iopscience.iop.org/article/10.1088/1367-2630/13/5/053035/meta}
  {\bibfield  {journal} {\bibinfo  {journal} {New Journal of Physics}\ }\textbf
  {\bibinfo {volume} {13}},\ \bibinfo {pages} {053035} (\bibinfo {year}
  {2011})}\BibitemShut {NoStop}%
\bibitem [{\citenamefont {Raghunandan}\ \emph {et~al.}(2018)\citenamefont
  {Raghunandan}, \citenamefont {Wrachtrup},\ and\ \citenamefont
  {Weimer}}]{raghunandan2018high}%
  \BibitemOpen
  \bibfield  {author} {\bibinfo {author} {\bibfnamefont {M.}~\bibnamefont
  {Raghunandan}}, \bibinfo {author} {\bibfnamefont {J.}~\bibnamefont
  {Wrachtrup}}, \ and\ \bibinfo {author} {\bibfnamefont {H.}~\bibnamefont
  {Weimer}},\ }\href
  {https://journals.aps.org/prl/abstract/10.1103/PhysRevLett.120.150501}
  {\bibfield  {journal} {\bibinfo  {journal} {Physical Review Letters}\
  }\textbf {\bibinfo {volume} {120}},\ \bibinfo {pages} {150501} (\bibinfo
  {year} {2018})}\BibitemShut {NoStop}%
\bibitem [{\citenamefont {Dicke}(1954)}]{dicke1954coherence}%
  \BibitemOpen
  \bibfield  {author} {\bibinfo {author} {\bibfnamefont {R.~H.}\ \bibnamefont
  {Dicke}},\ }\href
  {https://journals.aps.org/pr/abstract/10.1103/PhysRev.93.99} {\bibfield
  {journal} {\bibinfo  {journal} {Physical Review}\ }\textbf {\bibinfo {volume}
  {93}},\ \bibinfo {pages} {99} (\bibinfo {year} {1954})}\BibitemShut {NoStop}%
\bibitem [{\citenamefont {Zapasskii}(2013)}]{zapasskii2013spin}%
  \BibitemOpen
  \bibfield  {author} {\bibinfo {author} {\bibfnamefont {V.~S.}\ \bibnamefont
  {Zapasskii}},\ }\href
  {https://www.osapublishing.org/aop/abstract.cfm?uri=aop-5-2-131} {\bibfield
  {journal} {\bibinfo  {journal} {Advances in Optics and Photonics}\ }\textbf
  {\bibinfo {volume} {5}},\ \bibinfo {pages} {131} (\bibinfo {year}
  {2013})}\BibitemShut {NoStop}%
\bibitem [{\citenamefont {Pfeuty}(1970)}]{pfeuty1970one}%
  \BibitemOpen
  \bibfield  {author} {\bibinfo {author} {\bibfnamefont {P.}~\bibnamefont
  {Pfeuty}},\ }\href {https://www.math.ucdavis.edu/~bxn/pfeuty1970.pdf}
  {\bibfield  {journal} {\bibinfo  {journal} {ANNALS of Physics}\ }\textbf
  {\bibinfo {volume} {57}},\ \bibinfo {pages} {79} (\bibinfo {year}
  {1970})}\BibitemShut {NoStop}%
\bibitem [{\citenamefont {Yang}\ and\ \citenamefont
  {Jacob}()}]{yang2018numerical}%
  \BibitemOpen
  \bibfield  {author} {\bibinfo {author} {\bibfnamefont {L.-P.}\ \bibnamefont
  {Yang}}\ and\ \bibinfo {author} {\bibfnamefont {Z.}~\bibnamefont {Jacob}},\
  }\href@noop {} {\ }\bibinfo {note} {Under prepairation.}\BibitemShut {Stop}%
\bibitem [{\citenamefont {Zhang}\ \emph {et~al.}(2014)\citenamefont {Zhang},
  \citenamefont {Yu}, \citenamefont {Liang}, \citenamefont {Chen},
  \citenamefont {Jia},\ and\ \citenamefont {Nori}}]{zhang2014quantum}%
  \BibitemOpen
  \bibfield  {author} {\bibinfo {author} {\bibfnamefont {Y.}~\bibnamefont
  {Zhang}}, \bibinfo {author} {\bibfnamefont {L.}~\bibnamefont {Yu}}, \bibinfo
  {author} {\bibfnamefont {J.-Q.}\ \bibnamefont {Liang}}, \bibinfo {author}
  {\bibfnamefont {G.}~\bibnamefont {Chen}}, \bibinfo {author} {\bibfnamefont
  {S.}~\bibnamefont {Jia}}, \ and\ \bibinfo {author} {\bibfnamefont
  {F.}~\bibnamefont {Nori}},\ }\href
  {https://www.nature.com/articles/srep04083} {\bibfield  {journal} {\bibinfo
  {journal} {Scientific reports}\ }\textbf {\bibinfo {volume} {4}},\ \bibinfo
  {pages} {4083} (\bibinfo {year} {2014})}\BibitemShut {NoStop}%
\bibitem [{\citenamefont {Imry}(1980)}]{imry1980finite}%
  \BibitemOpen
  \bibfield  {author} {\bibinfo {author} {\bibfnamefont {Y.}~\bibnamefont
  {Imry}},\ }\href {https://journals.aps.org/prb/pdf/10.1103/PhysRevB.21.2042}
  {\bibfield  {journal} {\bibinfo  {journal} {Physical Review B}\ }\textbf
  {\bibinfo {volume} {21}},\ \bibinfo {pages} {2042} (\bibinfo {year}
  {1980})}\BibitemShut {NoStop}%
\bibitem [{\citenamefont {Skotiniotis}\ \emph {et~al.}(2015)\citenamefont
  {Skotiniotis}, \citenamefont {Sekatski},\ and\ \citenamefont
  {D{\"u}r}}]{skotiniotis2015quantum}%
  \BibitemOpen
  \bibfield  {author} {\bibinfo {author} {\bibfnamefont {M.}~\bibnamefont
  {Skotiniotis}}, \bibinfo {author} {\bibfnamefont {P.}~\bibnamefont
  {Sekatski}}, \ and\ \bibinfo {author} {\bibfnamefont {W.}~\bibnamefont
  {D{\"u}r}},\ }\href
  {http://iopscience.iop.org/article/10.1088/1367-2630/17/7/073032/meta}
  {\bibfield  {journal} {\bibinfo  {journal} {New Journal of Physics}\ }\textbf
  {\bibinfo {volume} {17}},\ \bibinfo {pages} {073032} (\bibinfo {year}
  {2015})}\BibitemShut {NoStop}%
\bibitem [{sup()}]{supplementary}%
  \BibitemOpen
  \href@noop {} {}\bibinfo {note} {See Supplemental Material for more
  information about the ground state wave function, the details of the
  application in quantum metrology, and the analogy of our proposed detector to
  the practical single-photon detectors.}\BibitemShut {Stop}%
\bibitem [{\citenamefont {Dziarmaga}(2010)}]{dziarmaga2010dynamics}%
  \BibitemOpen
  \bibfield  {author} {\bibinfo {author} {\bibfnamefont {J.}~\bibnamefont
  {Dziarmaga}},\ }\href
  {https://www.tandfonline.com/doi/abs/10.1080/00018732.2010.514702} {\bibfield
   {journal} {\bibinfo  {journal} {Advances in Physics}\ }\textbf {\bibinfo
  {volume} {59}},\ \bibinfo {pages} {1063} (\bibinfo {year}
  {2010})}\BibitemShut {NoStop}%
\bibitem [{\citenamefont {Yang}\ \emph {et~al.}(2018)\citenamefont {Yang},
  \citenamefont {Tang},\ and\ \citenamefont {Jacob}}]{yang2018concept}%
  \BibitemOpen
  \bibfield  {author} {\bibinfo {author} {\bibfnamefont {L.-P.}\ \bibnamefont
  {Yang}}, \bibinfo {author} {\bibfnamefont {H.~X.}\ \bibnamefont {Tang}}, \
  and\ \bibinfo {author} {\bibfnamefont {Z.}~\bibnamefont {Jacob}},\ }\href
  {https://journals.aps.org/pra/abstract/10.1103/PhysRevA.97.013833} {\bibfield
   {journal} {\bibinfo  {journal} {Physical Review A}\ }\textbf {\bibinfo
  {volume} {97}},\ \bibinfo {pages} {013833} (\bibinfo {year}
  {2018})}\BibitemShut {NoStop}%
\bibitem [{\citenamefont {Dziarmaga}(2005)}]{Dziarmaga2005Dynamics}%
  \BibitemOpen
  \bibfield  {author} {\bibinfo {author} {\bibfnamefont {J.}~\bibnamefont
  {Dziarmaga}},\ }\href {\doibase 10.1103/PhysRevLett.95.245701} {\bibfield
  {journal} {\bibinfo  {journal} {Phys. Rev. Lett.}\ }\textbf {\bibinfo
  {volume} {95}},\ \bibinfo {pages} {245701} (\bibinfo {year}
  {2005})}\BibitemShut {NoStop}%
\bibitem [{\citenamefont {Zurek}\ \emph {et~al.}(2005)\citenamefont {Zurek},
  \citenamefont {Dorner},\ and\ \citenamefont {Zoller}}]{Zurek2005Dynamics}%
  \BibitemOpen
  \bibfield  {author} {\bibinfo {author} {\bibfnamefont {W.~H.}\ \bibnamefont
  {Zurek}}, \bibinfo {author} {\bibfnamefont {U.}~\bibnamefont {Dorner}}, \
  and\ \bibinfo {author} {\bibfnamefont {P.}~\bibnamefont {Zoller}},\ }\href
  {\doibase 10.1103/PhysRevLett.95.105701} {\bibfield  {journal} {\bibinfo
  {journal} {Phys. Rev. Lett.}\ }\textbf {\bibinfo {volume} {95}},\ \bibinfo
  {pages} {105701} (\bibinfo {year} {2005})}\BibitemShut {NoStop}%
\bibitem [{\citenamefont {Heyl}\ \emph {et~al.}(2013)\citenamefont {Heyl},
  \citenamefont {Polkovnikov},\ and\ \citenamefont
  {Kehrein}}]{Heyl2013Dynamical}%
  \BibitemOpen
  \bibfield  {author} {\bibinfo {author} {\bibfnamefont {M.}~\bibnamefont
  {Heyl}}, \bibinfo {author} {\bibfnamefont {A.}~\bibnamefont {Polkovnikov}}, \
  and\ \bibinfo {author} {\bibfnamefont {S.}~\bibnamefont {Kehrein}},\ }\href
  {\doibase 10.1103/PhysRevLett.110.135704} {\bibfield  {journal} {\bibinfo
  {journal} {Phys. Rev. Lett.}\ }\textbf {\bibinfo {volume} {110}},\ \bibinfo
  {pages} {135704} (\bibinfo {year} {2013})}\BibitemShut {NoStop}%
\bibitem [{\citenamefont {Abeling}\ and\ \citenamefont
  {Kehrein}(2016)}]{Abeling2016quantum}%
  \BibitemOpen
  \bibfield  {author} {\bibinfo {author} {\bibfnamefont {N.~O.}\ \bibnamefont
  {Abeling}}\ and\ \bibinfo {author} {\bibfnamefont {S.}~\bibnamefont
  {Kehrein}},\ }\href {\doibase 10.1103/PhysRevB.93.104302} {\bibfield
  {journal} {\bibinfo  {journal} {Phys. Rev. B}\ }\textbf {\bibinfo {volume}
  {93}},\ \bibinfo {pages} {104302} (\bibinfo {year} {2016})}\BibitemShut
  {NoStop}%
\bibitem [{\citenamefont {\ifmmode \check{Z}\else
  \v{Z}\fi{}unkovi\ifmmode~\check{c}\else \v{c}\fi{}}\ \emph
  {et~al.}(2018)\citenamefont {\ifmmode \check{Z}\else
  \v{Z}\fi{}unkovi\ifmmode~\check{c}\else \v{c}\fi{}}, \citenamefont {Heyl},
  \citenamefont {Knap},\ and\ \citenamefont {Silva}}]{Zunkovic2018dynamical}%
  \BibitemOpen
  \bibfield  {author} {\bibinfo {author} {\bibfnamefont {B.}~\bibnamefont
  {\ifmmode \check{Z}\else \v{Z}\fi{}unkovi\ifmmode~\check{c}\else
  \v{c}\fi{}}}, \bibinfo {author} {\bibfnamefont {M.}~\bibnamefont {Heyl}},
  \bibinfo {author} {\bibfnamefont {M.}~\bibnamefont {Knap}}, \ and\ \bibinfo
  {author} {\bibfnamefont {A.}~\bibnamefont {Silva}},\ }\href {\doibase
  10.1103/PhysRevLett.120.130601} {\bibfield  {journal} {\bibinfo  {journal}
  {Phys. Rev. Lett.}\ }\textbf {\bibinfo {volume} {120}},\ \bibinfo {pages}
  {130601} (\bibinfo {year} {2018})}\BibitemShut {NoStop}%
\bibitem [{\citenamefont {Pelissetto}\ \emph {et~al.}(2018)\citenamefont
  {Pelissetto}, \citenamefont {Rossini},\ and\ \citenamefont
  {Vicari}}]{pelissetto2018out}%
  \BibitemOpen
  \bibfield  {author} {\bibinfo {author} {\bibfnamefont {A.}~\bibnamefont
  {Pelissetto}}, \bibinfo {author} {\bibfnamefont {D.}~\bibnamefont {Rossini}},
  \ and\ \bibinfo {author} {\bibfnamefont {E.}~\bibnamefont {Vicari}},\ }\href
  {\doibase 10.1103/PhysRevB.97.094414} {\bibfield  {journal} {\bibinfo
  {journal} {Phys. Rev. B}\ }\textbf {\bibinfo {volume} {97}},\ \bibinfo
  {pages} {094414} (\bibinfo {year} {2018})}\BibitemShut {NoStop}%
\bibitem [{\citenamefont {Heyl}(2018)}]{heyl2018dynamical}%
  \BibitemOpen
  \bibfield  {author} {\bibinfo {author} {\bibfnamefont {M.}~\bibnamefont
  {Heyl}},\ }\href
  {http://iopscience.iop.org/article/10.1088/1361-6633/aaaf9a/meta} {\bibfield
  {journal} {\bibinfo  {journal} {Reports on Progress in Physics}\ }\textbf
  {\bibinfo {volume} {81}},\ \bibinfo {pages} {054001} (\bibinfo {year}
  {2018})}\BibitemShut {NoStop}%
\bibitem [{\citenamefont {Yuen}(1976)}]{yuen1976states}%
  \BibitemOpen
  \bibfield  {author} {\bibinfo {author} {\bibfnamefont {H.}~\bibnamefont
  {Yuen}},\ }\href
  {https://ac.els-cdn.com/0375960176901602/1-s2.0-0375960176901602-main.pdf?_tid=1bc436de-d7d1-4600-98e4-372552700193&acdnat=1527015118_4044815a2f64d4db1a90d85e8d282365}
  {\bibfield  {journal} {\bibinfo  {journal} {Physics Letters A}\ }\textbf
  {\bibinfo {volume} {56}},\ \bibinfo {pages} {105} (\bibinfo {year}
  {1976})}\BibitemShut {NoStop}%
\bibitem [{\citenamefont {Baumann}\ \emph {et~al.}(2010)\citenamefont
  {Baumann}, \citenamefont {Guerlin}, \citenamefont {Brennecke},\ and\
  \citenamefont {Esslinger}}]{baumann2010dicke}%
  \BibitemOpen
  \bibfield  {author} {\bibinfo {author} {\bibfnamefont {K.}~\bibnamefont
  {Baumann}}, \bibinfo {author} {\bibfnamefont {C.}~\bibnamefont {Guerlin}},
  \bibinfo {author} {\bibfnamefont {F.}~\bibnamefont {Brennecke}}, \ and\
  \bibinfo {author} {\bibfnamefont {T.}~\bibnamefont {Esslinger}},\ }\href
  {https://www.nature.com/articles/nature09009} {\bibfield  {journal} {\bibinfo
   {journal} {Nature}\ }\textbf {\bibinfo {volume} {464}},\ \bibinfo {pages}
  {1301} (\bibinfo {year} {2010})}\BibitemShut {NoStop}%
\bibitem [{\citenamefont {Zhang}\ \emph {et~al.}(2017)\citenamefont {Zhang},
  \citenamefont {Pagano}, \citenamefont {Hess}, \citenamefont {Kyprianidis},
  \citenamefont {Becker}, \citenamefont {Kaplan}, \citenamefont {Gorshkov},
  \citenamefont {Gong},\ and\ \citenamefont {Monroe}}]{zhang2017observation}%
  \BibitemOpen
  \bibfield  {author} {\bibinfo {author} {\bibfnamefont {J.}~\bibnamefont
  {Zhang}}, \bibinfo {author} {\bibfnamefont {G.}~\bibnamefont {Pagano}},
  \bibinfo {author} {\bibfnamefont {P.~W.}\ \bibnamefont {Hess}}, \bibinfo
  {author} {\bibfnamefont {A.}~\bibnamefont {Kyprianidis}}, \bibinfo {author}
  {\bibfnamefont {P.}~\bibnamefont {Becker}}, \bibinfo {author} {\bibfnamefont
  {H.}~\bibnamefont {Kaplan}}, \bibinfo {author} {\bibfnamefont {A.~V.}\
  \bibnamefont {Gorshkov}}, \bibinfo {author} {\bibfnamefont {Z.-X.}\
  \bibnamefont {Gong}}, \ and\ \bibinfo {author} {\bibfnamefont
  {C.}~\bibnamefont {Monroe}},\ }\href
  {https://www.nature.com/articles/nature24654} {\bibfield  {journal} {\bibinfo
   {journal} {Nature}\ }\textbf {\bibinfo {volume} {551}},\ \bibinfo {pages}
  {601} (\bibinfo {year} {2017})}\BibitemShut {NoStop}%
\bibitem [{\citenamefont {Jurcevic}\ \emph {et~al.}(2017)\citenamefont
  {Jurcevic}, \citenamefont {Shen}, \citenamefont {Hauke}, \citenamefont
  {Maier}, \citenamefont {Brydges}, \citenamefont {Hempel}, \citenamefont
  {Lanyon}, \citenamefont {Heyl}, \citenamefont {Blatt},\ and\ \citenamefont
  {Roos}}]{Jurcevic2017direct}%
  \BibitemOpen
  \bibfield  {author} {\bibinfo {author} {\bibfnamefont {P.}~\bibnamefont
  {Jurcevic}}, \bibinfo {author} {\bibfnamefont {H.}~\bibnamefont {Shen}},
  \bibinfo {author} {\bibfnamefont {P.}~\bibnamefont {Hauke}}, \bibinfo
  {author} {\bibfnamefont {C.}~\bibnamefont {Maier}}, \bibinfo {author}
  {\bibfnamefont {T.}~\bibnamefont {Brydges}}, \bibinfo {author} {\bibfnamefont
  {C.}~\bibnamefont {Hempel}}, \bibinfo {author} {\bibfnamefont {B.~P.}\
  \bibnamefont {Lanyon}}, \bibinfo {author} {\bibfnamefont {M.}~\bibnamefont
  {Heyl}}, \bibinfo {author} {\bibfnamefont {R.}~\bibnamefont {Blatt}}, \ and\
  \bibinfo {author} {\bibfnamefont {C.~F.}\ \bibnamefont {Roos}},\ }\href
  {\doibase 10.1103/PhysRevLett.119.080501} {\bibfield  {journal} {\bibinfo
  {journal} {Phys. Rev. Lett.}\ }\textbf {\bibinfo {volume} {119}},\ \bibinfo
  {pages} {080501} (\bibinfo {year} {2017})}\BibitemShut {NoStop}%
\bibitem [{\citenamefont {Bernien}\ \emph {et~al.}(2017)\citenamefont
  {Bernien}, \citenamefont {Schwartz}, \citenamefont {Keesling}, \citenamefont
  {Levine}, \citenamefont {Omran}, \citenamefont {Pichler}, \citenamefont
  {Choi}, \citenamefont {Zibrov}, \citenamefont {Endres}, \citenamefont
  {Greiner} \emph {et~al.}}]{bernien2017probing}%
  \BibitemOpen
  \bibfield  {author} {\bibinfo {author} {\bibfnamefont {H.}~\bibnamefont
  {Bernien}}, \bibinfo {author} {\bibfnamefont {S.}~\bibnamefont {Schwartz}},
  \bibinfo {author} {\bibfnamefont {A.}~\bibnamefont {Keesling}}, \bibinfo
  {author} {\bibfnamefont {H.}~\bibnamefont {Levine}}, \bibinfo {author}
  {\bibfnamefont {A.}~\bibnamefont {Omran}}, \bibinfo {author} {\bibfnamefont
  {H.}~\bibnamefont {Pichler}}, \bibinfo {author} {\bibfnamefont
  {S.}~\bibnamefont {Choi}}, \bibinfo {author} {\bibfnamefont {A.~S.}\
  \bibnamefont {Zibrov}}, \bibinfo {author} {\bibfnamefont {M.}~\bibnamefont
  {Endres}}, \bibinfo {author} {\bibfnamefont {M.}~\bibnamefont {Greiner}},
  \emph {et~al.},\ }\href {https://www.nature.com/articles/nature24622}
  {\bibfield  {journal} {\bibinfo  {journal} {Nature}\ }\textbf {\bibinfo
  {volume} {551}},\ \bibinfo {pages} {579} (\bibinfo {year}
  {2017})}\BibitemShut {NoStop}%
\bibitem [{\citenamefont {Harris}\ \emph {et~al.}(2018)\citenamefont {Harris},
  \citenamefont {Sato}, \citenamefont {Berkley}, \citenamefont {Reis},
  \citenamefont {Altomare}, \citenamefont {Amin}, \citenamefont {Boothby},
  \citenamefont {Bunyk}, \citenamefont {Deng}, \citenamefont {Enderud},
  \citenamefont {Huang}, \citenamefont {Hoskinson}, \citenamefont {Johnson},
  \citenamefont {Ladizinsky}, \citenamefont {Ladizinsky}, \citenamefont
  {Lanting}, \citenamefont {Li}, \citenamefont {Medina}, \citenamefont
  {Molavi}, \citenamefont {Neufeld}, \citenamefont {Oh}, \citenamefont
  {Pavlov}, \citenamefont {Perminov}, \citenamefont {Poulin-Lamarre},
  \citenamefont {Rich}, \citenamefont {Smirnov}, \citenamefont {Swenson},
  \citenamefont {Tsai}, \citenamefont {Volkmann}, \citenamefont {Whittaker},\
  and\ \citenamefont {Yao}}]{Harris2018Phase}%
  \BibitemOpen
  \bibfield  {author} {\bibinfo {author} {\bibfnamefont {R.}~\bibnamefont
  {Harris}}, \bibinfo {author} {\bibfnamefont {Y.}~\bibnamefont {Sato}},
  \bibinfo {author} {\bibfnamefont {A.~J.}\ \bibnamefont {Berkley}}, \bibinfo
  {author} {\bibfnamefont {M.}~\bibnamefont {Reis}}, \bibinfo {author}
  {\bibfnamefont {F.}~\bibnamefont {Altomare}}, \bibinfo {author}
  {\bibfnamefont {M.~H.}\ \bibnamefont {Amin}}, \bibinfo {author}
  {\bibfnamefont {K.}~\bibnamefont {Boothby}}, \bibinfo {author} {\bibfnamefont
  {P.}~\bibnamefont {Bunyk}}, \bibinfo {author} {\bibfnamefont
  {C.}~\bibnamefont {Deng}}, \bibinfo {author} {\bibfnamefont {C.}~\bibnamefont
  {Enderud}}, \bibinfo {author} {\bibfnamefont {S.}~\bibnamefont {Huang}},
  \bibinfo {author} {\bibfnamefont {E.}~\bibnamefont {Hoskinson}}, \bibinfo
  {author} {\bibfnamefont {M.~W.}\ \bibnamefont {Johnson}}, \bibinfo {author}
  {\bibfnamefont {E.}~\bibnamefont {Ladizinsky}}, \bibinfo {author}
  {\bibfnamefont {N.}~\bibnamefont {Ladizinsky}}, \bibinfo {author}
  {\bibfnamefont {T.}~\bibnamefont {Lanting}}, \bibinfo {author} {\bibfnamefont
  {R.}~\bibnamefont {Li}}, \bibinfo {author} {\bibfnamefont {T.}~\bibnamefont
  {Medina}}, \bibinfo {author} {\bibfnamefont {R.}~\bibnamefont {Molavi}},
  \bibinfo {author} {\bibfnamefont {R.}~\bibnamefont {Neufeld}}, \bibinfo
  {author} {\bibfnamefont {T.}~\bibnamefont {Oh}}, \bibinfo {author}
  {\bibfnamefont {I.}~\bibnamefont {Pavlov}}, \bibinfo {author} {\bibfnamefont
  {I.}~\bibnamefont {Perminov}}, \bibinfo {author} {\bibfnamefont
  {G.}~\bibnamefont {Poulin-Lamarre}}, \bibinfo {author} {\bibfnamefont
  {C.}~\bibnamefont {Rich}}, \bibinfo {author} {\bibfnamefont {A.}~\bibnamefont
  {Smirnov}}, \bibinfo {author} {\bibfnamefont {L.}~\bibnamefont {Swenson}},
  \bibinfo {author} {\bibfnamefont {N.}~\bibnamefont {Tsai}}, \bibinfo {author}
  {\bibfnamefont {M.}~\bibnamefont {Volkmann}}, \bibinfo {author}
  {\bibfnamefont {J.}~\bibnamefont {Whittaker}}, \ and\ \bibinfo {author}
  {\bibfnamefont {J.}~\bibnamefont {Yao}},\ }\href
  {http://science.sciencemag.org/content/361/6398/162} {\bibfield  {journal}
  {\bibinfo  {journal} {Science}\ }\textbf {\bibinfo {volume} {361}},\ \bibinfo
  {pages} {162} (\bibinfo {year} {2018})}\BibitemShut {NoStop}%
\bibitem [{\citenamefont {Yang}\ \emph {et~al.}(2012)\citenamefont {Yang},
  \citenamefont {Li},\ and\ \citenamefont {Sun}}]{yang2012franck}%
  \BibitemOpen
  \bibfield  {author} {\bibinfo {author} {\bibfnamefont {L.-P.}\ \bibnamefont
  {Yang}}, \bibinfo {author} {\bibfnamefont {Y.}~\bibnamefont {Li}}, \ and\
  \bibinfo {author} {\bibfnamefont {C.}~\bibnamefont {Sun}},\ }\href
  {https://link.springer.com/article/10.1140/epjd/e2012-30324-9} {\bibfield
  {journal} {\bibinfo  {journal} {The European Physical Journal D}\ }\textbf
  {\bibinfo {volume} {66}},\ \bibinfo {pages} {300} (\bibinfo {year}
  {2012})}\BibitemShut {NoStop}%
\bibitem [{\citenamefont {Husimi}(1940)}]{husimi1940some}%
  \BibitemOpen
  \bibfield  {author} {\bibinfo {author} {\bibfnamefont {K.}~\bibnamefont
  {Husimi}},\ }\href
  {https://www.jstage.jst.go.jp/article/ppmsj1919/22/4/22_4_264/_pdf/-char/ja}
  {\bibfield  {journal} {\bibinfo  {journal} {Proceedings of the
  Physico-Mathematical Society of Japan. 3rd Series}\ }\textbf {\bibinfo
  {volume} {22}},\ \bibinfo {pages} {264} (\bibinfo {year} {1940})}\BibitemShut
  {NoStop}%
\bibitem [{\citenamefont {Lee}(1984)}]{Lee1984Qfunction}%
  \BibitemOpen
  \bibfield  {author} {\bibinfo {author} {\bibfnamefont {C.~T.}\ \bibnamefont
  {Lee}},\ }\href {\doibase 10.1103/PhysRevA.30.3308} {\bibfield  {journal}
  {\bibinfo  {journal} {Phys. Rev. A}\ }\textbf {\bibinfo {volume} {30}},\
  \bibinfo {pages} {3308} (\bibinfo {year} {1984})}\BibitemShut {NoStop}%
\bibitem [{\citenamefont {Radcliffe}(1971)}]{radcliffe1971some}%
  \BibitemOpen
  \bibfield  {author} {\bibinfo {author} {\bibfnamefont {J.}~\bibnamefont
  {Radcliffe}},\ }\href
  {http://iopscience.iop.org/article/10.1088/0305-4470/4/3/009/meta} {\bibfield
   {journal} {\bibinfo  {journal} {Journal of Physics A: General Physics}\
  }\textbf {\bibinfo {volume} {4}},\ \bibinfo {pages} {313} (\bibinfo {year}
  {1971})}\BibitemShut {NoStop}%
\bibitem [{\citenamefont {Arecchi}\ \emph {et~al.}(1972)\citenamefont
  {Arecchi}, \citenamefont {Courtens}, \citenamefont {Gilmore},\ and\
  \citenamefont {Thomas}}]{arecchi1972atomic}%
  \BibitemOpen
  \bibfield  {author} {\bibinfo {author} {\bibfnamefont {F.}~\bibnamefont
  {Arecchi}}, \bibinfo {author} {\bibfnamefont {E.}~\bibnamefont {Courtens}},
  \bibinfo {author} {\bibfnamefont {R.}~\bibnamefont {Gilmore}}, \ and\
  \bibinfo {author} {\bibfnamefont {H.}~\bibnamefont {Thomas}},\ }\href
  {https://journals.aps.org/pra/abstract/10.1103/PhysRevA.6.2211} {\bibfield
  {journal} {\bibinfo  {journal} {Physical Review A}\ }\textbf {\bibinfo
  {volume} {6}},\ \bibinfo {pages} {2211} (\bibinfo {year} {1972})}\BibitemShut
  {NoStop}%
\bibitem [{\citenamefont {Demkowicz-Dobrza{\'n}ski}\ \emph
  {et~al.}(2012)\citenamefont {Demkowicz-Dobrza{\'n}ski}, \citenamefont
  {Ko{\l}ody{\'n}ski},\ and\ \citenamefont
  {Gu{\c{t}}{\u{a}}}}]{demkowicz2012elusive}%
  \BibitemOpen
  \bibfield  {author} {\bibinfo {author} {\bibfnamefont {R.}~\bibnamefont
  {Demkowicz-Dobrza{\'n}ski}}, \bibinfo {author} {\bibfnamefont
  {J.}~\bibnamefont {Ko{\l}ody{\'n}ski}}, \ and\ \bibinfo {author}
  {\bibfnamefont {M.}~\bibnamefont {Gu{\c{t}}{\u{a}}}},\ }\href
  {https://www.nature.com/articles/ncomms2067} {\bibfield  {journal} {\bibinfo
  {journal} {Nature communications}\ }\textbf {\bibinfo {volume} {3}},\
  \bibinfo {pages} {1063} (\bibinfo {year} {2012})}\BibitemShut {NoStop}%
\bibitem [{\citenamefont {Lita}\ \emph {et~al.}(2008)\citenamefont {Lita},
  \citenamefont {Miller},\ and\ \citenamefont {Nam}}]{lita2008counting}%
  \BibitemOpen
  \bibfield  {author} {\bibinfo {author} {\bibfnamefont {A.~E.}\ \bibnamefont
  {Lita}}, \bibinfo {author} {\bibfnamefont {A.~J.}\ \bibnamefont {Miller}}, \
  and\ \bibinfo {author} {\bibfnamefont {S.~W.}\ \bibnamefont {Nam}},\ }\href
  {https://www.osapublishing.org/oe/abstract.cfm?uri=oe-16-5-3032} {\bibfield
  {journal} {\bibinfo  {journal} {Optics express}\ }\textbf {\bibinfo {volume}
  {16}},\ \bibinfo {pages} {3032} (\bibinfo {year} {2008})}\BibitemShut
  {NoStop}%
\bibitem [{\citenamefont {Korzh}\ \emph {et~al.}(2018)\citenamefont {Korzh},
  \citenamefont {Zhao}, \citenamefont {Frasca}, \citenamefont {Allmaras},
  \citenamefont {Autry}, \citenamefont {Bersin}, \citenamefont {Colangelo},
  \citenamefont {Crouch}, \citenamefont {Dane}, \citenamefont {Gerrits} \emph
  {et~al.}}]{korzh2018demonstrating}%
  \BibitemOpen
  \bibfield  {author} {\bibinfo {author} {\bibfnamefont {B.}~\bibnamefont
  {Korzh}}, \bibinfo {author} {\bibfnamefont {Q.}~\bibnamefont {Zhao}},
  \bibinfo {author} {\bibfnamefont {S.}~\bibnamefont {Frasca}}, \bibinfo
  {author} {\bibfnamefont {J.}~\bibnamefont {Allmaras}}, \bibinfo {author}
  {\bibfnamefont {T.}~\bibnamefont {Autry}}, \bibinfo {author} {\bibfnamefont
  {E.}~\bibnamefont {Bersin}}, \bibinfo {author} {\bibfnamefont
  {M.}~\bibnamefont {Colangelo}}, \bibinfo {author} {\bibfnamefont
  {G.}~\bibnamefont {Crouch}}, \bibinfo {author} {\bibfnamefont
  {A.}~\bibnamefont {Dane}}, \bibinfo {author} {\bibfnamefont {T.}~\bibnamefont
  {Gerrits}},  \emph {et~al.},\ }\href {https://arxiv.org/abs/1804.06839}
  {\bibfield  {journal} {\bibinfo  {journal} {arXiv preprint arXiv:1804.06839}\
  } (\bibinfo {year} {2018})}\BibitemShut {NoStop}%
\bibitem [{\citenamefont {Schuck}\ \emph {et~al.}(2013)\citenamefont {Schuck},
  \citenamefont {Pernice},\ and\ \citenamefont {Tang}}]{schuck2013waveguide}%
  \BibitemOpen
  \bibfield  {author} {\bibinfo {author} {\bibfnamefont {C.}~\bibnamefont
  {Schuck}}, \bibinfo {author} {\bibfnamefont {W.~H.}\ \bibnamefont {Pernice}},
  \ and\ \bibinfo {author} {\bibfnamefont {H.~X.}\ \bibnamefont {Tang}},\
  }\href {https://www.nature.com/articles/srep01893} {\bibfield  {journal}
  {\bibinfo  {journal} {Scientific Reports}\ }\textbf {\bibinfo {volume} {3}},\
  \bibinfo {pages} {1893} (\bibinfo {year} {2013})}\BibitemShut {NoStop}%
\end{thebibliography}%

\newpage
\appendix

\begin{widetext}

\part{Supplementary: Quantum Critical Detector---Amplifying Weak Signals Using First-Order Dynamical Quantum Phase Transitions}

\section{Defining quantum phases in the detector}
By defining the collective angular momentum operators of the $N$ spins 
\begin{equation}
\hat{S}_{\alpha}=\frac{1}{2}\sum_{j=1}^{N}\hat{\sigma}_{j}^{\alpha},\ \alpha=x,y,z,    
\end{equation}
we can rewrite our model Hamiltonian as
\begin{equation}
H=\hat{d}^{\dagger}\hat{d}+\frac{2\lambda}{\sqrt{N}}\hat{S}_{x}(\hat{d} + \hat{d}^{\dagger}) + \epsilon\hat{S}_{z} - \frac{2J}{N}\hat{S}_y^2.\!\label{eq:H_full}
\end{equation} 
One can see that the spin ensemble is equivalent to a single particle with spin-$N/2$~\cite{yang2012franck}. As the total angular momentum of the spin ensemble is conserved, we can perform the numerical simulation within a subspace spanned by the $N+1$ Dicke states~\cite{dicke1954coherence}. This makes our model become tractable in numerical simulation.

There exist three quantum phases in our model: paramagnetic-normal (PN) phase, ferromagnetic-normal (FN) phase, and ferromagnetic-superradiant (FS) phase. The phase diagram of our model can be obtained via direct numerical evaluation. By exploiting a mean-field theory~\cite{zhang2014quantum}, we verify our numerical simulation and also obtain the exact analytical wave function of the ground states~\cite{yang2018numerical}. A short summary of the phase boundaries, ground states, and the values of the OPs in each phase is given in table~\ref{tab:OrderParameters}. 

\section{Ground-State Wave Function of the detector}

To reveal the underlying microscopic mechanism of our quantum critical detector (QCD), we need to understand the fundamental changes in the ground state of the system during the quantum phase transitions (QPTs). As we known, the Husimi $Q$-function is a quantum analog of the classical distribution function in the phase space, which provide a powerful tool to vividly show the ground-state wave function of the system in each phase. 

\begin{table*}
\begin{tabular}{|c|c||c||c|c|c|}
\hline 
 & Stability Conditions & Ground States $|\sqrt{N}\alpha\rangle \otimes\left|\theta,\phi\right\rangle $ & $\zeta_{M,x}=\langle\hat{S}_{x}^{2}\rangle/N^{2}$ & $\zeta_{M,y}=\langle\hat{S}_{y}^{2}\rangle/N^{2}$ & $\zeta_{S}=\langle\hat{d}^{\dagger}\hat{d}\rangle/N$\tabularnewline
\hline 
PN Phase & $\epsilon>4\lambda^{2},\ \epsilon>2J$ & $\left|0\right\rangle \left|0,\phi_{0}\right\rangle $ & $0$ & $0$ & $0$\tabularnewline
\hline 
FN Phase & $J>\epsilon/2,\ J>2\lambda^{2}$ & $\left|0\right\rangle \left|\theta_{0},\phi_{0}\right\rangle ,\ \phi_{0}=\frac{\pi}{2},\frac{3\pi}{2}$  & $0$ & Finite & $0$\tabularnewline
\hline 
FS Phase & $4\lambda^{2}>\epsilon,\ 2\lambda^{2}>J$ & $|-\sqrt{N}\alpha_{0}e^{i\phi_{0}}\rangle\left|\theta_{0},\phi_{0}\right\rangle \!,\ \phi_{0}\!=\!0,\!\pi$  & Finite & $0$ & Finite \tabularnewline
\hline 
\end{tabular}\caption{\label{tab:OrderParameters}A summary of stability conditions, the
ground states, and the mean value of the order parameters in different
quantum phases. The phase boundaries are determined by the energy splitting of the spins $\epsilon$, the spin-boson coupling $\lambda$, and the spin-spin coupling $J$. State $|\sqrt{N}\alpha\rangle \otimes\left|\theta,\phi\right\rangle $
denotes the tensor product of the bosonic coherent state $|\sqrt{N}\alpha\rangle$ and a coherent
spin state $\left|\theta,\phi\right\rangle$. Ferromagnetic phases have two degenerate states with
different azimuth angle $\phi_{0}$ and the ground state in this phase can be any superposition of these two degenerate ground states.}
\end{table*}

\begin{figure}
\includegraphics[width=16cm]{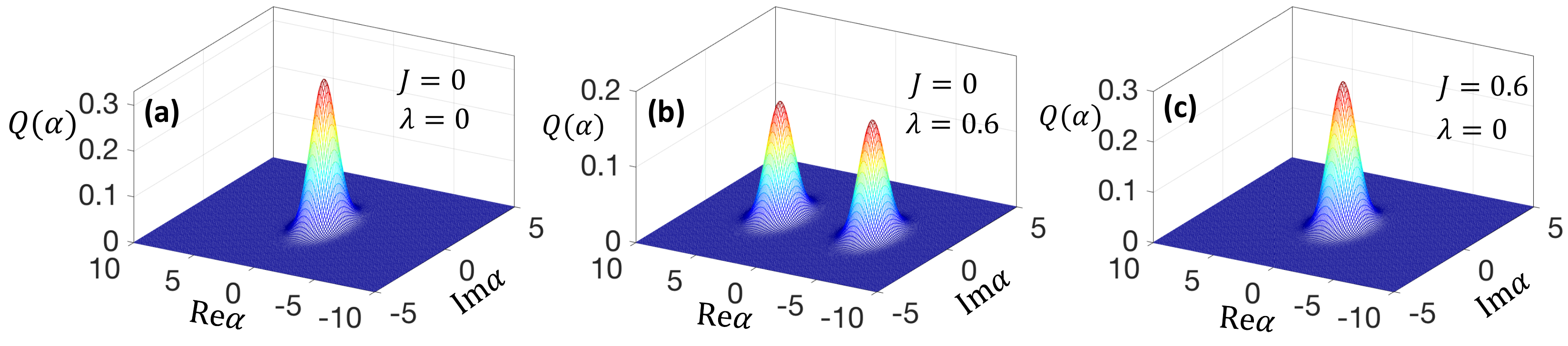}
\caption{\label{fig:QFboson} The Husimi $Q$-functions of the bosonic mode on the ground states of the paramagnetic-normal phase (a), the ferromagnetic-superradiant phase (b), and the ferromagnetic-normal phase (c) are displayed. Here, the other parameters are set as $\epsilon=1$, spin number $N=80$, and the bosonic mode cutoff $80$.}
\end{figure}

The bosonic Husimi $Q$ is defined as~\cite{husimi1940some},
\begin{equation}
Q(\alpha) = \frac{1}{\pi}\rm{Tr}_{\rm spin}[\langle\alpha|\rho_g|\alpha\rangle], \label{eq:QF_boson}
\end{equation}
where $\rho_g$ is the density matrix of the ground state, $|\alpha \rangle$ is an arbitrary bosonic coherent state, and $\rm{Tr}_{\rm spin}[\cdots]$ means tracing off the degrees of freedom of the spins. For the spin degree of freedom, the Bloch sphere is usually utilized to characterize an arbitrary state of spin-$1/2$ particle. But for a spin-$n$ particle, a $[(2n+1)^2-1]$-sphere is required. It is impossible to show such a high-dimensional sphere. To overcome this issue, we  introduce the spin Husimi $Q$-function~\cite{Lee1984Qfunction},
\begin{equation}
Q(\theta,\phi)=\frac{2N+1}{4\pi}\rm{Tr}_{\rm boson}[\left\langle \theta,\phi\right|\rho_g\left|\theta,\phi\right\rangle],\label{eq:spinQF}
\end{equation}
where $N$ is the spin number, $\rm{Tr}_{\rm boson}[\cdots]$ means tracing off the degrees of freedom of the bosonic mode, and $|\theta,\phi\rangle$ is a coherent spin state~\cite{radcliffe1971some,arecchi1972atomic}. 
To let the angle $\theta$ be the exact same polar angle of a spherical coordinate, we redefine the coherent spin state as,
\begin{equation}
\left|\theta,\phi\right\rangle=e^{i\theta(\hat{S}_{x}\sin\phi-\hat{S}_{y}\cos\phi)}\left|N/2,N/2\right\rangle.   
\end{equation}
Here, $|N/2,N/2\rangle$ is the Dicke state with all the spins on the up state.

\begin{figure}
\includegraphics[width=16cm]{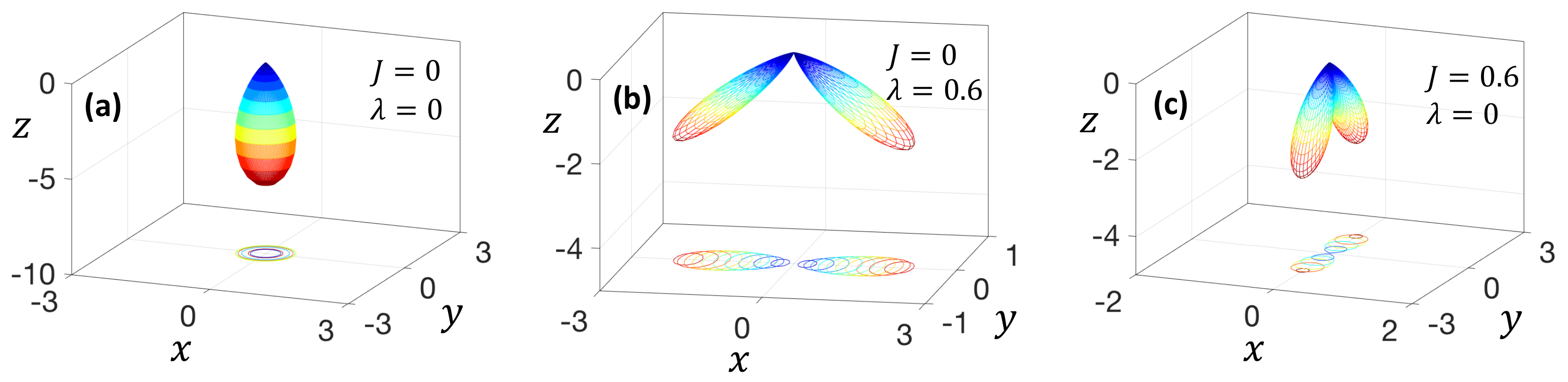}
\caption{\label{fig:QFSpin} The Husimi $Q$-functions of the spins for the ground states $\rho_g$ of the paramagnetic-normal phase (a), the ferromagnetic-superradiant phase (b), and the ferromagnetic-normal phase (c) are displayed. Here, the spherical coordinates $(r=Q(\theta,\phi),\theta,\phi)$ have been transferred to the corresponding Cartesian coordinates $(x,y,z)$. The curves underneath are the contour projections of the corresponding $Q$-functions in $xy$-plane. The other parameters are set as $\epsilon=1$, spin number $N=80$, and the bosonic mode cutoff $80$.}
\end{figure}

Our numerical approach allows us to directly calculate the ground states and the corresponding Husimi $Q$-functions of the system. In the paramagnetic-normal (PN) phase, the bosonic mode is on the vacuum state. As shown in Fig.~\ref{fig:QFboson} (a), the bosonic $Q$-function has only one peak located at the origin. In this phase, the spins are on the Dicke state $|N/2,-N/2\rangle$, i.e., all the spins are polarized along negative $z$-axis. As shown in Fig.~\ref{fig:QFSpin} (a), the corresponding spin $Q$-function is a cigar-like structure lying along negative $z$-axis. From this quasi-probability distribution function, we can see that the mean magnetization $M_z=\langle\hat{S}_z\rangle_0/N$ ($\langle\cdots\rangle_0$ means averaging on the ground state) is a finite negative value, but the two magnetic order parameters $\zeta_{M,x}=\langle\hat{S}^2_x\rangle_0/N^2$ and $\zeta_{M,y}=\langle\hat{S}^2_y\rangle_0/N^2$ are very small and will go to zero in the limit $N\rightarrow\infty$.

The ferromagnetic-superradiant (FS) phase has two degenerate ground states $|\alpha_0\rangle\otimes |\theta_0,\pi\rangle$ and $|-\alpha_0\rangle\otimes |\theta_0,0\rangle$. Here, $|\alpha_0\rangle$ is a bosonic coherent state and $|\theta_0,\phi_0\rangle$ ($\phi_0=0,\ \pi$) is a coherent spin state. The value of $\alpha_0$ and $\theta_0$ can be determined by the mean-field theory in the thermodynamic limit $N\rightarrow\infty$~\cite{yang2018numerical}.  When the
system adiabatically goes to the FS phase, the system can be on an arbitrary superposition of these two degenerate states. Thus, the ensemble mean value of the displacement of the bosonic mode $\langle\hat{d}^{\dagger}+\hat{d}\rangle_0$, the magnetization along $x$-direction $\langle\hat{S}_{x}\rangle_{0}$, and the magnetization along $y$-direction $\langle\hat{S}_{y}\rangle_{0}$ are all zero. But the excitation number reflected in the superradiant order parameter $\zeta_{S}=\langle\hat{d}^{\dagger}\hat{d}\rangle_0/N$ and the magnetic noise characterized by the magnetic order parameter $\zeta_{M,x}$ are finite. In the numerical simulation, we choose the ground state as a symmetric superposition of these two degenerate states. In Fig.~\ref{fig:QFboson} (b), we show the $Q$-function of the bosonic mode. The two separated peaks on the real axis indicate the macroscopic excitation in the bosonic mode. In Fig.~\ref{fig:QFSpin} (b), we also see the rotation of the spins in the $xz$-plane as the strong spin-boson coupling $\lambda$ is along $x$-axis. These two branches correspond to the two degenerate states.

The ferromagnetic-normal (FN) phase also has two degenerate states $|0\rangle\otimes|\theta_0,\pi/2\rangle$ and $|0\rangle\otimes|\theta_0,3\pi/2\rangle$. When the system adiabatically goes to the FN phase, the bosonic mode is always on the vacuum state as shown by the corresponding $Q$-function in Fig.~\ref{fig:QFboson} (c). But the spins can be on an arbitrary superposition of these two coherent spin states $|\theta_0,\phi_0\rangle$ ($\phi_0=\pi/2,\ 3\pi/3$). Thus, the ensemble mean of the magnetization in $xy$-plane is still zero. But the magnetic noise characterized by the magnetic order parameter $\zeta_{M,x}$ is finite. The $Q$-function of the spin in the FN phase is displayed in Fig.~\ref{fig:QFSpin} (c). The strong spin-spin coupling $J$ along $y$-axis leads to the rotation of the spins in $yz$-plane. Here, we can also see that, when QPT from the FN phase to the FS phase occurs, the energy prestored in the spins transfers to the bosonic mode contributing to the macroscopic excitation in the bosonic mode. Also, a fundamental change in the spin noise from the $y$-direction to the $x$-direction can be observed. 

The QPTs of PN$\leftrightarrow$FN and PN$\leftrightarrow$FS are of second order. Only the transition FN$\leftrightarrow$FS is a first order one. We emphasize that second order phase transitions which are continuous phase transitions do not possess the necessary features for quantum critical detection. In stark contrast, first order discontinuous phase transitions exhibit a giant response when a weak perturbation is applied making them an ideal resource for quantum critical detection.

\section{Beating the Heisenberg limit with first-order quantum phase transitions}
Our first-order quantum phase transition model has two main applications: (1) quantum parameter estimation (time-independent process with $N^2$ scaling); (2) quantum dynamical amplification (time-dependent process with $N$ scaling as explained in the next section). For applications such as quantum metrology, we do not use phase measurements or interferometry as in the conventional parameter estimation. However, we would like to emphasize the striking fact that we can beat the Heisenberg limit using quantum criticality in first-order phase transition as explain in the following. 

We first give a short review of the Heisenberg limit. In the traditional interference measurements, like the Mach-Zehnder interference and the atomic Ramsey interference measurements, the $1/N$ scaling lays a general upper bound to the rate of the parameter estimation error decreasing with the source (e.g., energy, number of probes) used in the measurement. This $1/N$ scaling is also called Heisenberg limit in quantum metrology~\cite{giovannetti2004quantum,giovannetti2006quantum}. The Heisenberg limit is essentially based on two main prerequisites: (1) the output signal is a continuous function of the parameter to be estimated in those measurements, like in the Mach-Zehnder interferometer, the intensity of the output light and the phase difference of the two paths is connected by $I_{\rm out}(\phi )\propto \sin^2\phi$ as shown in Fig.~\ref{fig:Metrology} (a). Thus, using the error propagation relation, one can easily connect the uncertainty of the parameter and the uncertainty of the output signal with $\Delta \phi=[\partial I_{\rm out}(\phi)/\partial \phi]^{-1}\Delta I_{\rm out}$; (2) The probes do not interact with each other. Thus, the accumulated dynamical phase of $N$ synchronized probes in a entangled state is $N$ time larger than that of single probe, like the dynamical phase of $N$ identical two-level atoms in the GHZ state $[\exp(-iN \phi)|e_1e_2\cdots e_N\rangle + |g_1g_2\cdots g_N\rangle]/\sqrt{2}$ ($\phi=\omega t$). For an ideal noiseless quantum setup, the Heisenberg limit can be reached with the GHZ-like state~\cite{demkowicz2012elusive}.   

\begin{figure}
\includegraphics[width=10cm]{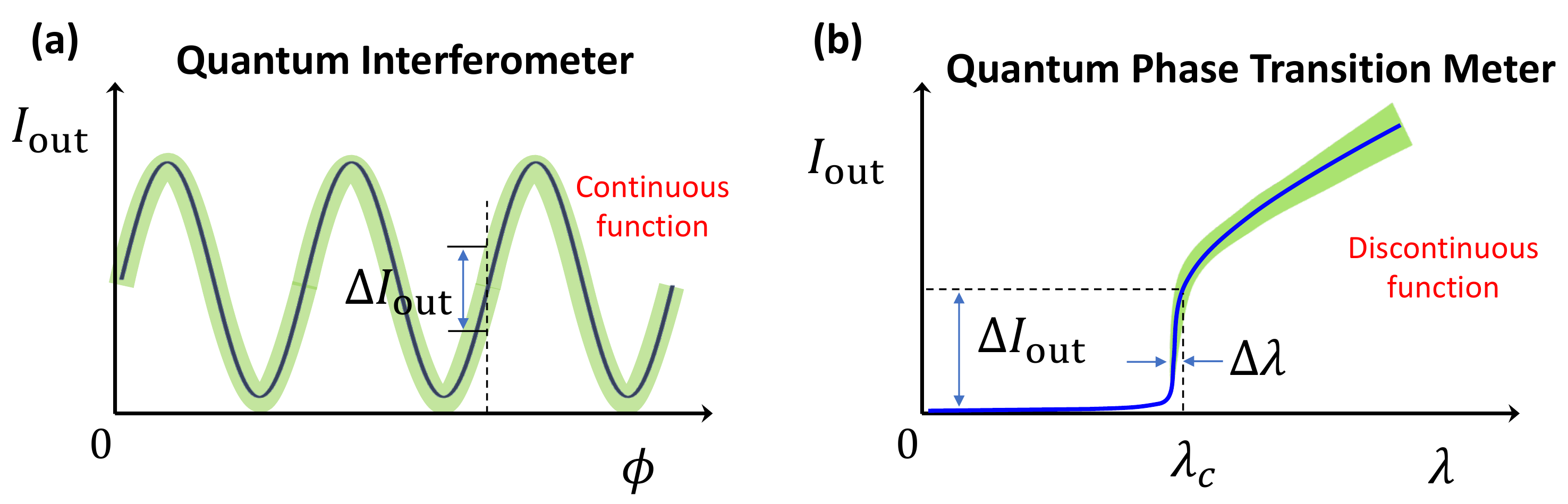}
\caption{\label{fig:Metrology} Schematic of estimation error in quantum metrology using a quantum interferometer (a) and a first-order quantum phase transition meter (b). In panel (a), the output of the interferometer is a continuous function of the phase to be estimated. The estimation error is connected to the uncertainty of the output signal~$\Delta \phi = [\partial I_{\rm out}(\phi)/\partial\phi]^{-1}\Delta I_{\rm out}$. In panel (b), both the mean value and the uncertainty of the output signal are discontinuous functions of the system parameter. Thus, the estimation error is limited by the width of the rising window $\Delta \lambda = \chi^{-1}(\lambda_c) \Delta I_{\rm out}$, which is inversely proportional to the sensitivity function (the slope at the critical point) $\chi (\lambda_c)\propto N^2$ ($N$ is the spin number).}
\end{figure}

Now, we explain how this Heisenberg limit can be beaten with first-order QPTs. During first-order QPTs, both of these two preconditions mentioned above have been broken. First, both the mean value and the uncertainty of the output signal are discontinuous functions of the system parameter at the critical point. In the FN phase, both the mean value and the uncertainty of the excitation number in the output bosonic mode are zero in the thermodynamic limit. Thus, at the critical point, an abrupt change both in the mean value and the quantum uncertainty of the output signal occurs. After the phase transition to the FS phase, the output signal is large enough to be distinguished from the FN phase. In this case, the estimation error is not limited by uncertainty of the output signal, but the width of the signal rising window at the critical point $\Delta \lambda = \chi^{-1}(\lambda_c) \Delta I_{\rm out}$ [see Fig.~\ref{fig:Metrology} (b)]. Here, $\Delta I_{\rm out}$ is the difference (the jump) of the order parameter in two neighboring quantum phases at the boundary, which has been re-scaled by the spin number $N$ as in the order parameter $\zeta_S=\langle \hat{d}^{\dagger}\hat{d}\rangle_0/N $. The function $\chi(\lambda)$ is the slope of the order parameter, which is defined as the sensitivity function in this paper.  Second, for our $N$-interacting-spin system, we have shown that $\chi$ diverges with the spin number in $N^2$ scaling at the first-order QPT critical point. Thus, this $N^{2}$ scaling can be used to enhance the sensitivity and beat the Heisenberg limit in parameter estimation. 

We also emphasize that a dynamical process is not necessarily required for metrological applications. In parameter estimation, each time we can precisely set the spin-boson coupling strength and cool down the system to its ground state. After that, a measurement can be performed on the excitation number in the bosonic model.  Then we change the spin-boson coupling strength and repeat this measurement. Finally, we can find the critical spin-boson coupling strength $\lambda_{c}$. Thus, the  scaling in first-order QPT can be used to beat the Heisenberg limit. Please note we use the spin-boson coupling strength only as an example to elucidate the scheme but the spin-spin coupling (J) can also be estimated with a similar measurement scheme. We can also use the highly-controllable spin-boson coupling strength to estimate the homogeneous spin-spin coupling with the Dicke-LMGy model. The estimation error of the spin-spin coupling is given by $\Delta J_y=2\lambda_{c}\Delta \lambda\propto 1/N^2$.

\section{Analogy with The Practical Single-photon Detectors}
As we mentioned in the main text, there are two main amplification schemes: quantum linear amplifiers and critically biased amplifiers. We are focusing on the second amplification scheme, where the weak input signal does not get amplified directly. Instead, it functions as a control of an optimally biased critical system, which is significantly different from the first one, such as quantum linear amplifiers. In these critically biased amplifiers, the input and output information carriers can be fundamentally different (eg: input photons and output electrons) and the corresponding gain is defined as the ratio of the outputs with and without the input control signal. In this section, we show the analogy of our proposed QCD with the practical critical detectors to explain the motivation of our work more clearly. Finally, we show that first-order QPT-based devices can pave a way for new types of weak signal detector. 

We first take the superconducting nanowire single-photon detector (SSNPD) as an example to explain the critical amplification scheme explicitly.  The SSNPD is the best available and widely used near-infrared single-photon detector with $\geq 95\%$ quantum efficiency~\cite{lita2008counting}, $<3$ picoseconds timing jitter~\cite{korzh2018demonstrating} and $<1$ dark count per hour~\cite{schuck2013waveguide}. The input is a single-photon pulse---an extremely weak quantum signal. The current in the superconducting nanowire is biased very close to the critical current, thus even a single-photon pulse can break the superconductivity. The output signal is the voltage difference between the two ends of the superconducting nanowire. In the transduction (absorption) process, the incident single-photon pulse will general one resonantly excited electron. As the center frequency of the pulse is much larger than the energy gap of the superconductor, this highly excited electron will break hundreds of Cooper pairs, reduce the local density of the superconducting electrons, and finally triggers a phase transition from a superconductor to a normal metal. Before the absorption of the photon, the voltage difference is extremely small. After the absorption of the photon, the superconducting nanowire becomes a normal metal. An observable output voltage pulse will be generated to realize the amplification. This process can be modeled as a time-varying local temperature induced thermodynamic phase transition. During the amplification process, the exact dynamical change in the superconducting electron density and the effective time-dependent local temperature can be obtained by numerical simulation.

This type of critically biased amplifiers has been extensively used in practical measurements even outside the context of SNSPDs, like the photomultiplier tube, the single-photon avalanche diode, single-electron transistor, etc. However, our proposed QCD is the first quantum analog of the classical critical detectors. Similarly, we also need to bias the detector very close to the critical point. After the absorption of the input weak signal, a time-dependent variation in the system parameter (such as the spin-boson coupling or spin-spin interaction strength) is induced to trigger a first-order dynamical QPT. For a specific measurement process, we also need to calculate the exact form of time-dependent change in system parameters. Without loss of generality, we  assume the time-varying spin-boson coupling change is proportional to the signal absorption probability $P_e(t)$ as shown in the subplot of Fig.~4 in the main text. We show that,  if the QCD is biased close to the critical point, a large amplification factor (the quantum gain) can be obtained. 

We emphasize that this weak-signal detection is significantly different from the parameter estimation process. During the dynamical amplification, the system parameter (coupling strength) is varied across the phase boundary time-dependently by the incident pulse. It is widely debated and is an open question whether QPTs retain their criticality during a dynamical process. The discontinuous change of the observable and the $N^2$ sensitivity found in the first-order QPT only occurs if the system goes from the ground state in one phase to the ground state of another phase. However, in a dynamical process, the system will not go to the ground state of the other phase but evolves to some complicated excited state. Therefore, we numerically studied the dynamics of the system around the critical point. We showed for the first time that the high sensitivity also exists in a dynamical process and thus explicitly demonstrated the dynamical amplification.

\end{widetext}
\end{document}